\documentclass[12pt]{article}

\usepackage{graphicx}

\def\simlt{\stackrel{<}{{}_\sim}}
\def\simgt{\stackrel{>}{{}_\sim}}

\begin{document}
\DeclareGraphicsExtensions{.jpg,.pdf,.mps,.png}
\begin{flushright}
\baselineskip=12pt
EFI-08-01 \\
ANL-HEP-PR-08-11
\end{flushright}

\begin{center}
\vglue 1.5cm

{\Large\bf  Dynamically Solving the $\mu/B_\mu$ Problem in
Gauge-mediated Supersymmetry Breaking} \vglue 2.0cm {\Large Tao
Liu$^{a,}$\footnote{Email: taoliu@theory.uchicago.edu} and Carlos
E.M. Wagner$^{a,b,c,}$\footnote{Email: cwagner@hep.anl.gov}}
\vglue 1cm {$^a$ Enrico Fermi Institute and
$^b$ Kavli Institute for Cosmological Physics, \\
University of Chicago, 5640 S. Ellis Ave., Chicago, IL
60637\\\vglue 0.2cm
$^c$ HEP Division, Argonne National Laboratory,
9700 Cass Ave., Argonne, IL 60439

}
\end{center}

\vglue 1.0cm
\begin{abstract}

We provide a simple solution to the $\mu/B_\mu$ problem in the
gauge-mediated Next-to-Minimal Supersymmetric Standard Model. In
this model the messenger sector contains one pair of $3+\bar 3$
and one pair of $2+\bar 2$ messengers. These two messenger pairs
couple to different gauge singlets in the hidden sector in which
supersymmetry (SUSY) is broken. Such a gauge-mediation structure
can naturally arise in many backgrounds. Because of the two
effective SUSY breaking scales $\frac{\langle F_i\rangle}{\langle
M_i\rangle}$ in the messenger sector, the renormalization group
evolutions of the soft SUSY breaking parameters can be properly
modified, leading to a negative enough singlet soft mass square
$m_N^2(\Lambda_{EW})$ and hence reasonable $\mu/B_\mu$ values. In
most of the perturbative (up to the GUT scale) parameter region,
as a result, the electroweak scale is stabilized and
phenomenologically interesting mass spectra of particles and
superparticles are obtained. In addition, this model favors large
values of $\tan\beta$: $5 \sim 50$ and a heavy scalar spectrum.
With the relatively large $\tan\beta$, the light $U(1)_R$
pseudoscalar (mainly appearing in the low-scale gauge-mediated
SUSY breaking models) becomes extremely singlet-like, and is no
longer a problem in this model. These features apply to all cases
of low-, intermediate- and high-scale gauge-mediated SUSY
breaking.

\end{abstract}


\newpage

\section{Introduction}

The Standard Model (SM) provides an excellent description of all
particle physics interactions, excluding gravity. The excellent
agreement of the SM predictions with the measured precision
electroweak (EW) physics observables would be recovered in any
extension of the SM in which the new physics decouples fast from
physics at the weak scale. Supersymmetry (SUSY) is an example of
such an extension. Supersymmetric particles receive contribution
to their masses through gauge invariant operators which are
independent of the Higgs mechanism, and their effects decouple
fast as these masses are increased. Since these contributions to
the slepton and squark  masses are not necessarily alligned in
flavor space with the lepton and quark masses, new flavor
violating contributions become significant, leading to a potential
conflict with flavor physics observables.

In gauge mediated SUSY breaking models, SUSY breaking masses are
flavor independent at the messenger scale, leading to flavor
violating effects that are still controlled by the
Cabbibo-Kobayashi-Maskawa (CKM) matrix elements, enabling the
existence of superparticles with EW scale masses. Problems remain
in the Higgs sector, however, related to the origin and natural
relation between the Higgsino mass parameter $\mu$ and the Higgs
mixing mass term $B_{\mu}$.

In the minimal supersymmetric standard model(MSSM), we need a term
of the form
\begin{eqnarray}
\Delta \mathcal{L}=\int d^2\theta \mu \mathbf{H_d}\mathbf{H_u} +
h.c. \label{101}
\end{eqnarray}
to give the Higgsinos a mass. If $\mu\gg \Lambda_{EW}$, the Higgs
scalars in the chiral superfields obtain a large mass term in the
potential and the EW symmetry may not be broken. If $\mu\ll
\Lambda_{EW}$, the lightest chargino mass is lighter than
$m_W^2/M_2$ with $M_2$ being the soft mass of bino and winos, and
the experimental bounds can not be satisfied. Therefore we must
have
\begin{eqnarray}
\mu\sim \Lambda_{EW}. \label{102}
\end{eqnarray}
However, it is hard to understand why the $\mu$ parameter is of this
scale instead of the more fundamental Planck scale $M_P$,
considering that it is not related in any direct way to the SUSY
breaking sector of the MSSM. Introducing a dynamical mechanism may help
solve this so-called $\mu$ problem, but generally at the price
of introducing some new problems, e.g., $\mu/B_\mu$ problem. The
$\mu/B_\mu$ problem is related to the origin of the scalar soft
SUSY breaking Higgs mixing  mass term,
\begin{eqnarray}
\Delta V=B_\mu H_d H_u + h.c. \label{103}
\end{eqnarray}
To stabilize the EW scale $M_{EW}$, it is necessary to have
\begin{eqnarray}
B_\mu \sim \Lambda_{EW}^2 \sim \mu^2. \label{104}
\end{eqnarray}
In the context of a dynamical generation of $\mu$, however, it
is difficult to generate a $B_{\mu}$ satisfying this
relation. This is particularly true for the case of low-scale
gauge-mediated SUSY breaking.

So far, there are three main mechanisms to solve the $\mu/B_\mu$
problem. The first one is the Giudice-Masiero mechanism which is
the  first proposed to solve this problem in the context of
gravity-mediated SUSY breaking~\cite{Giudice:1988yz}. Its basic
idea is to assume an exact Peccei-Quinn symmetry,  forbidding the
$\mu$ term in the supersymmetric limit, and then generate it and
$B_\mu$ dynamically according to the SUSY breaking effects of the
same order. Explicitly, the authors of Ref.~\cite{Giudice:1988yz}
introduce one set of higher-dimensional operators in the Kahler
potential
\begin{eqnarray}
\Delta \mathcal{L}=\int d^4\theta
\mathbf{H_d}\mathbf{H_u}\Big(\frac{c_1}{M_P} \mathbf{X^\dagger} +
\frac{c_2}{M_P^2} \mathbf{X^\dagger}\mathbf{X}\Big) +h.c.
\label{105}
\end{eqnarray}
here $\mathbf{X}$ is the SUSY breaking chiral spurion. Once the
SUSY is broken, the effective $\mu$ and $B_\mu$ parameters are generated
as
\begin{eqnarray}
\mu = \frac{c_1 \langle F_X\rangle}{M_P}, \ \ \ \ \ \ B_\mu =
\frac{c_2 \langle F_X\rangle^2}{M_P^2}.
\end{eqnarray}
Since $\frac{ \langle F_X\rangle}{M_P}$ denotes the natural scale
of the soft terms in the gravity-mediation case, the correct
relationship
\begin{eqnarray}
\frac{B_\mu}{\mu} \sim \frac{c_2}{c_1}\frac{\langle F_X
\rangle}{M_P} \sim {\rm TeV} \label{106}
\end{eqnarray}
arises if $c_1\sim c_2 \sim \mathcal{O}(1)$. But this idea is hard
to be translated to the gauge-mediation case. In the effective
theory of gauge-mediation, the $\mu$ and $B_\mu$ operators
(similar to those in Eq.(\ref{105})) are generally induced at the
same loop-level. Unlike the gravity-mediation case, the effective
SUSY breaking scale $\frac{ \langle F_X\rangle}{\Lambda_M}$ is no
longer the natural scale of the soft terms, which necessarily
leads to a modified relationship between $\mu$ and $B_\mu$
\begin{eqnarray}
\frac{B_\mu}{\mu} \sim \frac{c_2}{c_1}\frac{\langle F_X
\rangle}{\Lambda_M}\sim \frac{\langle F_X \rangle}{\Lambda_M} \sim
100 {\rm TeV}. \label{108}
\end{eqnarray}
Here $\Lambda_M$ denotes the messenger scale (unlike the
gravity-mediation, $c_1$ and $c_2$ now represent the product of
coupling constants and possible loop factors).  Recently, it was
noticed~\cite{Roy:2007nz} that the $B_\mu$ operator in
Eq.(\ref{105}) is not protected by non-renormalization theorems of
the hidden sector because $\mathbf{X}^\dagger \mathbf{X}$ is not a
holomorphic or anti-holomorphic operator of the hidden sector. The
strong dynamics in the hidden sector therefore can efficiently
suppress $c_2$ with respect to $c_1$, in the renormalization
group(RG) evolution above the SUSY breaking scale $\sqrt{\langle
F_X\rangle}$. However, due to the same effect, the characteristic
mass spectrum of gauge mediation in the squark and slepton sectors
is ruined in this model. It turns out that the physically allowed
parameter region for this model is rather small~\cite{Cho:2008fr}.

A second one is the dynamical relaxation
mechanism~\cite{Dvali:1996cu}. Its basic idea is to generate $\mu$
and $B_\mu$ according to the SUSY breaking effects of different
orders. Explicitly, one can forbid the appearance of
non-holomorphic operators and hence a $B_\mu$ operator in the
effective action of one-loop level. Then the operators from the
higher-order corrections can be responsible of generating $B_\mu$
of the correct size. Such an one-loop effective action has the
general form~\cite{Giudice:2007ca}
\begin{eqnarray}
\Delta \mathcal{L}=\int d^4\theta
\mathbf{H_d}\mathbf{H_u}[f(\mathbf{X}) + g(\mathbf{X}^\dagger) +
D^2h(\mathbf{X},\mathbf{X}^\dagger)] +h.c. \label{109}
\end{eqnarray}
with $D_\alpha$ being the supersymmetric covariant derivative and
$f$, $g$, $h$ being generic functions of SUSY breaking chiral
spurion $\mathbf{X}$. These one-loop effective operators can be
induced by a proper construction of the superpotential in the
messenger sector. The effective $\mu$ term then arise according to
the second term~\cite{Giudice:2007ca} or the third
one~\cite{Dvali:1996cu} which are characterized by a divergent
logarithmic form of $\mathbf{X}^\dagger\mathbf{X}$ in this
mechanism. As for the $B_\mu$ term, it will be generated at a
higher order in perturbation theory. This mechanism is similar to
that of the soft mass generation of squarks and sleptons. But,
compared to the naturalness of the latter due to the absence of
couplings between the squarks, sleptons and the messenger sector
at tree level, the structure of the required superpotential for
the former is typically non-generic. Actually, a new dimensional
parameter is introduced again in the superpotential of the
messenger sector~\cite{Dvali:1996cu,Giudice:2007ca}.

The third one is the light singlet mechanism which is also the
focus of this paper. In this scenario, the $\mu$ term is forbidden
by some discrete symmetries (e.g., in the Next-to-Minimal
Supersymmetric Standard Model (NMSSM) and the nearly-Minimal
Supersymmetric Standard Model (nMSSM)) or by some additional
Abelian gauge symmetry (e.g., in the $U(1)'$-extended Minimal
Supersymmetric Standard Model (UMSSM)~\cite{Langacker:1999hs}).
With the introduction of a singlet chiral superfield $\mathbf{N}$
in the observable sector which has the coupling
\begin{eqnarray}
\Delta \mathcal{L}=\int d^2\theta \lambda
\mathbf{N}\mathbf{H_d}\mathbf{H_u} +h.c. \label{110}
\end{eqnarray}
the effective $\mu$ and $B_\mu$ parameters arise as
\begin{eqnarray}
\mu\equiv \lambda v_N, \ \ \ \ \ \
 B_\mu  \sim
\lambda \langle F_N \rangle, \label{111}
\end{eqnarray}
after the SUSY is broken. So, as long as the scalar $N$ and its
auxiliary field $F_N$ are stabilized at the soft SUSY breaking
scale or EW scale, a correct relationship
\begin{eqnarray}
\frac{B_\mu}{\mu} \sim \frac{\langle F_N\rangle}{v_N} \sim
10^2-10^3 {\rm GeV}. \label{112}
\end{eqnarray}
can be achieved.
But this mechanism faces serious problems: (1) to generate a
proper $v_N$, a negative enough soft mass square
$m_N^2(\Lambda_{EW})$ is required, which turns out to be a rather
difficult mission, persisting for any messenger
scale~\cite{Dine:1993yw,deGouvea:1997cx}; (2) both the trilinear
soft parameters $|A_\lambda(\Lambda_{EW})|$ and
$|A_\kappa(\Lambda_{EW})|$ are generically small, compared to
$\Lambda_{EW}$ (this is due to the fact that they are highly
suppressed at the messenger scale while their RG evolutions down
to the EW scale are mediated by small beta functions.). Since they
are the only sources explicitly breaking the global $U(1)_R$
symmetry, their smallness necessarily leads to a light
pseudoscalar which, unless is mainly a singlet, is ruled out by
the current LEP bound~\cite{Dine:1993yw}. One way to circumvent
these difficulties is to make the $\mathbf{N}$ field couple to the
messengers~\cite{Giudice:1997ni}(also see \cite{Han:1999jc}), or
to extra light freedom
degrees~\cite{Dine:1993yw,Langacker:1999hs}. Then a modified
boundary value (at the messenger scale) or beta function of
$m_N^2$ may help solve this problem. But, it was realized recently
that
the experimental bounds on the Higgs mass can add severe
constraints on the former class of models~\cite{Delgado:2007rz}.
As for the latter, it is viable, except that the couplings
generally need to be strong $\sim \mathcal{O}(1)$ if only small
number of the light freedom degrees exist~\cite{Langacker:1999hs}.
In this paper, however, we will show that the problems in the
light singlet mechanism are just some misguided images. In the
context of a more general gauge-mediated NMSSM where the (minimal)
messenger pairs $3+\bar 3$ and $2+\bar 2$ couple to different SUSY
breaking chiral spurions in the hidden sector, there is no
difficulty in generating a negative enough $m_N^2(\Lambda_{EW})$
in most of the perturbative (up to GUT scale) $\lambda-\kappa$
parameter region. The EW scale is then stabilized, and
phenomenologically interesting mass spectra of particles and
superparticles are also obtained. As a general feature, squarks
and sleptons become heavy, while there are light charginos and
neutralinos, which are mostly an admixture of Higgsinos and
singlinos. Such an interesting gauge-mediation structure can
effectively arise in many general backgrounds.

In addition, there is no light $U(1)_R$ pseudoscalar problem in our
model. For the intermediate- and high-scale gauge mediations,
large $|A_\lambda(\Lambda_{EW})|$ comparable with $\Lambda_{EW}$
are typical, so the lightest Higgs pseudoscalar actually are not
light. For the low-scale case, even though
$|A_\lambda(\Lambda_{EW})|$ and $|A_\kappa(\Lambda_{EW})|$ are not
always large, the lightest Higgs pseudoscalar is extremely
singlet-like due to large $\tan\beta$ values, escaping the
experimental constraints again. The light singlet mechanism
therefore is naturally implemented, without introducing any
complicated or special elements in the messenger sector. Our idea
is proposed in section 2, and followed are the numerical results
in section 3. The last section is our discussions and conclusions.
Since different energy scales are involved for the parameters in
this paper, we will specify them unless they can be understood
according to the context.

\section{A Simple Model to Solve $\mu/B\mu$ Problem}

\subsection{The NMSSM}

As the simplest extension of the MSSM, the NMSSM has a
superpotential for the Higgs superfields
\begin{eqnarray}
\mathbf{W}=\lambda \mathbf{N}\mathbf{H_d}\mathbf{H_u} -
\frac{1}{3}\kappa \mathbf{N}^3, \label{201}
\end{eqnarray}
where the $\mu$ term in the MSSM has been forbidden by a $Z_3$
discrete symmetry. The cubic term of $\mathbf{N}$ in the
superpotential explicitly breaks the Peccei-Quinn
symmetry
\begin{eqnarray}
\mathbf{N} \rightarrow \mathbf{N} e^{i \alpha}, \mathbf{H_d}
\mathbf{H_u} \rightarrow \mathbf{H_d} \mathbf{H_u}  e^{-i \alpha}.
\label{205}
\end{eqnarray}
In the absence of the singlet cubic term
the EW symmetry breaking would spontaneously break the
Peccei-Quinn symmetry as well, and hence lead to a dangerous
Peccei-Quinn Goldstone boson.


It is not hard to write down the tree-level neutral Higgs
potential in the NMSSM, which consists of $F$-terms, $D$-terms,
and soft SUSY-breaking terms
\begin{eqnarray}
V_0 & =& V_F+V_D+V_{soft}, \nonumber\\
 V_F &=& |\lambda H_d H_u - k N^2|^2 + \lambda ^2 |N|^2
(|H_d|^2+|H_u|^2),
           \nonumber  \\
 V_D  &=& \frac{g_Y^2+g_2^2}{8} (|H_d|^2-|H_u|^2)^2, \nonumber \\
 V_{\it soft} &=& {m^2_{H_d}} |H_d|^2 + {m^2_{H_u}} |H_u|^2
+{m^2_N}|N|^2 \nonumber \\
 && -
 (\lambda A_{\lambda} H_d H_u N + h.c.)-\left(\frac{\kappa}{3} A_{\kappa} N^3 +
h.c.\right). \label{202}
\end{eqnarray}
Here $H_d$, $H_u$ and $N$ denote the neutral Higgs bosons
corresponding to $\mathbf{H_d}$, $\mathbf{H_u}$ and $\mathbf{N}$,
respectively.

The one-loop effective Higgs potential is formally given by
\begin{equation}
\Delta V = \frac{1}{64\pi^2} \mbox{STr} {\cal M}^4 (H_i) \left(\ln
\frac{{\cal M}^2 (H_i)}{\Lambda_{\overline{\rm MS}}^2} -
\frac{3}{2} \right) \label{203}
\end{equation}
Here $ {\cal M}^2 (H_i)$ is a field-dependent mass-squared matrix,
and $\Lambda_{\overline{\rm MS}}$ is the $\overline{\rm MS}$
renormalization scale at which all
RG evoluved parameers are fixed. Since $\Delta V$ may bring significant
radiative corrections to some of the Higgs boson masses, we will
include it into our analysis. Two-loop corrections to the Higgs
potential will not be included, but we will comment on their possible
effects at the end of this article.

Note, even though the one-loop
effective Higgs potential brings an explicit dependence on
$\Lambda_{\overline{\rm MS}}$, all observables are independent of it.
The mass matrix $ {\cal M}^2$ depends on the fields $H_i$ through
their couplings to various other particles. Since it is the
strength of these couplings, instead of the absolute values of the
masses, that measures the one-loop corrections to the minimization
conditions and to the Higgs mass matrix, the most important
one-loop corrections come from the field-dependent masses of top
quark sector and bottom quark sector (for large $\tan\beta$ case).
In this paper, we will only consider their contributions to the
one-loop effective Higgs potential. Explicitly, they are given by
\begin{eqnarray}
\Delta V &=& \frac{3}{32\pi^2} \left[ m_{\tilde{t}_1}^4 (H_i)
\left( \ln \frac{m_{\tilde{t}_1}^2 (H_i)}{\Lambda_{\overline{\rm
MS}}^2} - \frac{3}{2} \right) + m_{\tilde{t}_2}^4 (H_i) \left( \ln
\frac{m_{\tilde{t}_2}^2 (H_i)}{\Lambda_{\overline{\rm MS}}^2} -
\frac{3}{2} \right) \right. \nonumber \\ && \left.  - 2 m_t^4
(H_i) \left( \ln \frac{m_t^2 (H_i)}{\Lambda_{\overline{\rm MS}}^2}
- \frac{3}{2} \right) \right] \nonumber
\\ && + \frac{3}{32\pi^2} \left[ m_{\tilde{b}_1}^4 (H_i)
\left( \ln \frac{m_{\tilde{b}_1}^2 (H_i)}{\Lambda_{\overline{\rm
MS}}^2} - \frac{3}{2} \right) + m_{\tilde{b}_2}^4 (H_i) \left( \ln
\frac{m_{\tilde{b}_2}^2 (H_i)}{\Lambda_{\overline{\rm MS}}^2} -
\frac{3}{2} \right) \right. \nonumber \\ && \left.  - 2 m_b^4
(H_i) \left( \ln \frac{m_b^2 (H_i)}{\Lambda_{\overline{\rm MS}}^2}
- \frac{3}{2} \right) \right]. \label{204}
\end{eqnarray}

\subsection{The Model}

\label{the model}

The NMSSM provides the simplest or most direct realization of the
light singlet mechanism to solve the $\mu/B_\mu$ problem in
gauge-mediated SUSY breaking models, where the $\mu/B_\mu$
parameters are effectively generated as
\begin{eqnarray}
\mu&\equiv &\lambda v_N, \nonumber \\
 B_\mu & \equiv &
\lambda \langle F_N \rangle +A_\lambda \mu, \label{301}
\end{eqnarray}
here
\begin{eqnarray}
v_N&=&\langle N \rangle \nonumber \\
\langle F_N \rangle &=& \frac{\kappa}{\lambda^2}\,\mu^2
-\frac{\lambda\, v^2}{2}\,\sin 2\beta \label{302}
\end{eqnarray}
with $v^2=v_d^2+ v_u^2$. But it is also confronted by the common
problems of all models of the light singlet mechanism. Let us
rephrase these problems in the framework of the NMSSM. Consider
the minimization conditions of the tree-level Higgs potential in
the NMSSM~\cite{deGouvea:1997cx}
\begin{eqnarray}
 \mu^2 &=& \lambda^2 v_N^2 =
   - \frac{M_Z^2}{2} + \frac{m_{H_d}^2-m_{H_u}^2 \tan^2 \beta}{\tan^2
   \beta-1}, \label{303}\\
    B_\mu &=&A_{\lambda} \lambda v_N - \lambda (\lambda v_d v_u-\kappa v_N^2) = (m_{H_d}^2+m_{H_u}^2+2 \lambda^2 v_N^2)
 \frac{\sin 2 \beta}{2} ,\label{304} \\
 2\kappa^2v_N^2 &=& \lambda v^2(\kappa\sin2\beta- \lambda) -
m_N^2 + A_{\lambda} \lambda v^2{\sin2\beta \over {2 v_N}} + \kappa
A_\kappa v_N . \label{305}
\end{eqnarray}
For the minimal gauge mediation, where the messenger sector is
\begin{eqnarray}
W= \lambda \mathbf{S} \bar \mathbf{q} \mathbf{q} + \gamma
\mathbf{S} \bar \mathbf{l} \mathbf{l} \label{306}
\end{eqnarray}
with $\mathbf{S}=S+\theta^2 F_S$ being the SUSY breaking spurion
field and $(\mathbf{q}+\mathbf{l})+(\bar \mathbf{q}+\bar
\mathbf{l})$ being $(3+2)+(\bar 3 + \bar 2) $ messenger pairs. The
lower mass bounds of sleptons or gluinos give a lower bound on the
effective SUSY breaking scale
\begin{eqnarray}
\Lambda_S=\frac{\langle F_S \rangle}{\langle S\rangle}
\end{eqnarray}
according to their RG evolutions. This lower bound immediately
implies another lower bound on the beta function of $m_{H_u}^2$
(because of its positivity), and then leads to
$m_{H_u}^2(\Lambda_{EW})\simlt -(200 {\rm
GeV})^2$~\cite{deGouvea:1997cx}. According to Eq.(\ref{303}), this
is translated into a stringent bound on the effective $\mu$
parameter or $v_N$ for mildly large
$\tan\beta$~\cite{deGouvea:1997cx}
\begin{eqnarray}
|\mu| \simgt 200{\rm GeV}. \label{307}
\end{eqnarray}
On the other hand, due to the third minimization condition
Eq.(\ref{305}) a very negative $m_N^2(\Lambda_{EW})$ or very large
$A_\lambda(\Lambda_{EW})$ and $A_\kappa(\Lambda_{EW})$ are
necessary in order to generate a large $v_N$ or $\mu$ for
$\lambda, \kappa \sim \mathcal{O}(1)$. This is difficult to
achieve
because (see the related RG equations summarized in Appendix A): \\
(1) For $m_N^2$, as the RG evolution runs down, its beta function
becomes small quickly due to the negative contribution from
$m_{H_u}^2$;  \\
(2) For $A_\lambda$ and $A_\kappa$, their values at the messenger
scale are highly suppressed due to their high-loop level origin
and, at the same time, their beta functions are not negative
enough. In addition, because $A_\lambda$ and $A_\kappa$ are the
only sources explicitly breaking the global $U(1)_R$ symmetry in
the Higgs potential, their smallness necessarily leads to an
almost massless pseudoscalar which is ruled out by the
current LEP bound~\cite{Dine:1993yw}.  \\
As a result, the $\mu/B_\mu$
problem is not solved in the NMSSM within the minimal gauge
mediation scenario~\cite{deGouvea:1997cx}.

In this paper, we present a new way to solve the $\mu/B_\mu$
problem within the NMSSM with gauge-mediated SUSY breaking.
Actually, we only take a simple modification to the
superpotential, Eq.(\ref{306}), assuming the new one to be
\begin{eqnarray}
W= \lambda \mathbf{S}_q \bar \mathbf{q} \mathbf{q} + \gamma
\mathbf{S}_l \bar \mathbf{l} \mathbf{l}, \label{308}
\end{eqnarray}
here $\mathbf{S}_q=S_q+\theta^2 F_q$ and
$\mathbf{S}_l=S_l+\theta^2F_l$ are two SUSY breaking chiral
spurions. In the following, we will use $\Lambda_q$ and
$\Lambda_l$ to denote the effective SUSY breaking scales, $i.e.$,
\begin{eqnarray}
\Lambda_q = \frac{\langle F_q\rangle}{\langle S_q\rangle}, \ \ \ \
\ \ \Lambda_l=\frac{\langle F_l\rangle}{\langle S_l\rangle}.
\label{319}
\end{eqnarray}
As explained above, the difficulty in generating a very negative
$m_N^2(\Lambda_{EW})$ is from the fact that the RG evolutions of
$m_{H_u}^2$ and $m_N^2$ are strongly coupled to each other. The
parameter $m_{H_u}^2$ has a large, positive beta function, so it
becomes negative quickly as the RG evolutions run down. The
negative $m_{H_u}^2$ leads to a negative contribution to the beta
function of $m_N^2$, therefore, preventing the appearance of a
large, negative $m_N^2(\Lambda_{EW})$. However, the story can be
dramatically changed after the introduction of a new parameter
$\eta=\Lambda_l/\Lambda_q$. In the minimal gauge mediation limit,
we have $\eta=1$. As it is increased, the beta function of $m_N^2$
is effectively enlarged according to the dominant terms
$4\lambda^2(m_{H_d}^2+m_{H_u}^2)$, while the beta function of
$m_{H_u}^2$ is effectively diminished according to the terms $-
(2g_Y^2M_1^2+6g_2^2M_2^2)$ (even though some other terms may have
positive contributions to this beta function.). Due to these
effects, the velocity for $m_N^2$ to evolve to a negative value is
increased, but that for $m_{H_u}^2$, is slowed down. It becomes
possible now to get a large, negative $m_N^2(\Lambda_{EW})$, even
if only a mild increase is made for $\eta$. In contrast to the
``minimal gauge mediation'', we will refer to this mechanism as
``general gauge mediation'' in the following.

With the superpotential of the messenger sector modified, the soft
SUSY breaking masses are also different from those generated in
the minimal gauge mediation case. These new soft masses at
the messenger scale are found to
be (see, for istance, Ref.~\cite{Wagner:1998vd})
\begin{eqnarray}
M_3 =  {\alpha_3 \over 4 \pi} \Lambda_q~~~~ M_2 = {\alpha_2 \over
4 \pi} \Lambda_l~~~~ M_1 = {\alpha_1 \over 4 \pi} \left
[\frac{2}{5} \Lambda_q + \frac{3}{5}\Lambda_l \right ] \label{309}
\end{eqnarray}
for gauginos, and
\begin{eqnarray}
m_\phi^2 =2 \left[ C_3^\phi\left({\alpha_3 \over 4 \pi}\right)^2
\Lambda_q^2 +C_2^\phi\left({\alpha_2\over 4 \pi}\right)^2
\Lambda_l^2 +C_1^\phi \left({\alpha_1\over 4 \pi}\right)^2
\left({2 \over 5} \Lambda_q^2 + {3 \over 5}\Lambda_l^2\right)
\right], \label{310}
\end{eqnarray}
for squarks, sleptons and neutral Higgs bosons. $C_3^\phi$,
$C_2^\phi$ and $C_1^\phi=\frac{3}{5} Y_\phi^2$ are quadratic
Casimir operators of the scalar $\phi$. It is easy to check that,
with $\Lambda_q=\Lambda_l$, Eq.(\ref{309}) and Eq.(\ref{310})
reduce to the results in the minimal gauge mediation case.

The superpotential of the general gauge mediation Eq~(\ref{308})
can naturally arise in many backgrounds. For example, in the case
where the messenger pairs $3+\bar 3$ and $2+\bar 2$ are coupled to
several SUSY breaking chiral spurions $\mathbf{S}_i=S_i+ \theta^2
F_i$ (e.g., see~\cite{Dine:2007dz})
\begin{eqnarray}
W= \lambda_i \mathbf{S}_i \bar \mathbf{q} \mathbf{q} + \gamma_i
\mathbf{S}_i \bar \mathbf{l} \mathbf{l}, \label{311}
\end{eqnarray}
we can assume
\begin{eqnarray}
\mathbf{S}_q = \lambda_i \mathbf{S}_i, \ \ \ \  \mathbf{S}_l =
\gamma_i \mathbf{S}_i.
\end{eqnarray}
$\Lambda_q$ and $\Lambda_l$ then effectively arise as
\begin{eqnarray}
\Lambda_q = {\lambda_i \langle F_i \rangle \over \lambda_j \langle
S_j \rangle}, \ \ \ \ \ \ \Lambda_l = {\gamma_i \langle F_i
\rangle \over \gamma_j \langle S_j\rangle}. \label{313}
\end{eqnarray}
Here the sums over the index "i" and "j" are implicitly assumed.

At last, it is necessary to point out that, with the one-loop
corrections to the Higgs potential included, the constraint on the
effective $\mu$ parameter given by (\ref{307}) can be relaxed to
some extent. As an illustration, let us consider the minimization
conditions with the corrections from the stops $\tilde t_1$ and
$\tilde t_2$ included
\begin{eqnarray}
 \mu^2 &=&
   - \frac{M_Z^2}{2} + \frac{m_{H_d}^2-m_{H_u}^2 \tan^2 \beta}{\tan^2
   \beta-1}
   + \frac{h_t^2 \sin^2\beta}{\cos 2\beta}(X_1+X_2) +{\mathcal O}(h_t\lambda,G^2), \label{314}\\
 B_\mu &=& (m_{H_d}^2+m_{H_u}^2+2 \lambda^2
 v_N^2)
 \frac{\sin 2 \beta}{2} + \frac{1}{2}h_t^2\sin2\beta(X_1+X_2)  +{\mathcal O}(h_t\lambda,G^2),\label{315} \\
 2\kappa^2v_N^2 &=& \lambda v^2(\kappa\sin2\beta- \lambda) -
m_N^2 + A_{\lambda} \lambda v^2{\sin2\beta \over {2 v_N}} + \kappa
A_\kappa v_N +{\mathcal O}(h_t\lambda,G^2). \label{316}
\end{eqnarray}
here
\begin{eqnarray}
G^2&=&g_Y^2+g_2^2,\label{317}\\ X_i&=&\frac{3}{32\pi^2}
 \left[2\left( \ln \frac{m_{\tilde{t}_i}^2 }{\Lambda_{\overline
{\rm MS}}^2} - 1 \right)m_{\tilde{t}_i}^2 \right] \label{318}
\end{eqnarray}
with $i=1,2$. Typically we have
\begin{eqnarray}
m_{\tilde t_{1,2}}^2(\Lambda_{\overline{\rm MS}}) \gg
\Lambda_{\overline {\rm MS}}^2
\end{eqnarray}
if $\Lambda_{\overline{\rm MS}} \sim m_t$ is assumed, which leads
to
\begin{eqnarray}
X_1+X_2 &\sim& 10^{-2} (m_{\tilde Q_{3}}^2(\Lambda_{\overline {\rm
MS}})+
m_{\tilde t}^2(\Lambda_{\overline {\rm MS}})) \nonumber \\
 &\sim& (100 {\rm GeV})^2,
\end{eqnarray}
for the soft masses $m_{\tilde Q_{3}}^2(\Lambda_{\overline {\rm
MS}})$ and $m_{\tilde t}^2(\Lambda_{\overline {\rm MS}})$ of
(TeV)$^2$ order. Given $\frac{h_t^2\sin^2\beta}{\cos 2\beta}<0$
for $\tan\beta>1$, the dominant effect of
$\frac{m_{H_d}^2-m_{H_u}^2 \tan^2 \beta}{\tan^2
   \beta-1}$ in mediating $\mu^2$ therefore
is weaken by the one-loop corrections. It turns out that, for an
effective $\mu=\lambda v_N$ as small as $100 {\rm GeV}$, the EW
scale can still be stabilized and phenomenologically interesting
physics can still arise (See Tables
(\ref{table1})-(\ref{table8})). More results from the numerical
analysis will be given in the next section.

\section{Numerical Analysis}

The general gauge mediation contains four unkown input parameters:
the superpotential couplings $\lambda(\Lambda_{EW})$ and
$\kappa(\Lambda_{EW})$, the messenger scale $\Lambda_M$ and the
ratio of the two effective SUSY breaking scales
$\eta=\Lambda_l/\Lambda_q$. All soft SUSY-breaking parameters at
the EW scale can be obtained by solving the RG equations
summarized in the Appendix A, with the boundary conditions at the
messenger scale $\Lambda_M$ given by Eqs.(\ref{309})-(\ref{310}).
As for the Yukawa couplings $h_t$ and $h_b$, even though we need
to give them initial values while minimizing the Higgs potential,
these values must be consistent with the masses of the top and
bottom quarks or the output values of $v_d$ and $v_u$\footnote{In
the numerical work, we use the tree level relationship
\begin{eqnarray}
h_t\approx\frac{165{\rm GeV}}{v_u}, \ \ \ \
h_b\approx\frac{h_t\tan\beta}{55}
\end{eqnarray}
where we have identified the running top-quark mass by applying
the appropriate QCD corrections to the top quark pole mass, and
the running mass of $m_b$ at the EW scale has been taken to be
about 3 GeV.}. So they are not true input parameters.


The introduction of the new parameter $\eta$ can lead to several
different phases after EW symmetry breaking: \\
(A) For fixed $\lambda$, $\kappa$ and $\Lambda_M$, if choose
$\eta\sim 1$, we recover the GFM phase discussed
in~\cite{deGouvea:1997cx}. As explained above, this phase does not
generate correct physics consistent
with the current experimental bounds~\cite{deGouvea:1997cx}.  \\
(B) With a further increased $\eta$, a new kind of phase with $v_d=v_N=0$ and
$v_u\neq 0$ may appear. Actually, this phase has been noticed in a
different background~\cite{Delgado:2007rz}.  \\
(C) We will refer to the third kind of phase as ``physical
$\mu/B_\mu$ phase''.
As $\eta$ increases, $v_N$ becomes large
compared to $ v=\sqrt {v_d^2+ v_u^2}=174 {\rm GeV}$. As a result,
one can always find an $\eta$ window characterized by $ v_N \gg v$
where the phenomenologically interesting physics can be generated.
Such $\eta$ windows will be our focus in this paper. Explicitly,
we will study such physical $\eta$ windows corresponding to
different points in the parameter space expanded by the other
three input parameters: $\lambda$, $\kappa$ and $\Lambda_M$. \\
(D) If $\eta$ keeps increasing, we will meet the last phase which
is characterized by $v_N\neq 0$ and $v_d=v_d=0$. This is caused by
large, negative $m_N^2(\Lambda_{EW})$.
In this case, the RG evolution of $m_{H_u}^2$ to negative values
is highly suppressed by the negative $m_N^2$ as well as the masses
of the EW gauginos. \\
The appearing of the multiple phases reflects the large freedom
degree caught by the parameter $\eta$. In the following we will
focus on the physics in the physical $\eta$ windows.

Our numerical results are summarized in
Tables(\ref{table1})-(\ref{table8}) which correspond to three
typical cases in the phase (C) discussed above: low-scale
($\Lambda_M\sim 10^5-10^6 {\rm GeV}$), intermediate-scale
($\Lambda_M\sim 10^{11} {\rm GeV}$) and high-scale ($\Lambda_M
\sim 10^{15} {\rm GeV}$) general gauge mediation\footnote{Some
recent papers point out that~\cite{Jedamzik:2005ir}: in the
context of $SU(5)$ gauge mediation, the requirement of a light
gravitino as dark matter favors the intermediate-scale scenario.
In this paper we will not expand this issue, but leave the
associated discussions to future work. }. We choose nine points on
the $\lambda(\Lambda_{EW})-\kappa(\Lambda_{EW})$ plane, and then
study their physics in all of the three cases which in turn helps
us figure out the related $\eta$ windows. The numerical results
show that to obtain consistency with current phenomenological
bounds, relatively small values of $\eta$, $\eta\sim 2$, are
required in the low-scale general gauge mediation. This should be
compared with values of $\eta \sim 4$ for the intermediate-scale
case and $\eta \sim 5$ for the high-scale case.

This can be simply
understood in the following way. As the path length of the RG
evolutions increases (due to the increase of $\Lambda_M$),
$m_{H_u}^2$ runs towards negative values at the late stage of its
evolution. This  makes the beta function of $m_N^2$ very small
or even negative. As a result, one cannot obtain large
enough negative values of $m_N^2(\Lambda_{EW})$. A larger $\eta$ can
help solve this problem, since it implies a smaller beta
function for $m_{H_u}^2$,
preventing $m_{H_u}^2$ from becoming
negative too fast.

Actually, the present experimental bounds lead to a more
complicated than the simple picture presented above.
In this model, the main constraints are from the
lightest chargino mass or the lightest $CP$-even Higgs mass,
depending on the messenger scale $\Lambda_M$. For
$M_2(\Lambda_{EW})\gg \mu > m_W$ (as typically happens in this
model), the lightest chargino mass is given by
\begin{eqnarray}
m_{\chi_1^c} = \mu + {\mathcal
O}\left(\frac{\mu}{M_2},\frac{m_W}{M_2}\right). \label{403}
\end{eqnarray}
As for the lightest $CP$-even Higgs, it is typically $H_u$-like
(because of the small mixing due to a not large
$A_\lambda(\Lambda_{EW})$ soft term). Its mass square at the tree
level is known to be typically less than $m_Z^2$, so the one-loop
corrections need to be included to escape the experimental bound.
In all the models we analyzed, the would be MSSM $CP$-odd Higgs
boson becomes very heavy and the lightest $CP$-even Higgs mass is
affected by a potentially large mixing with the $CP$-even singlet
state. As we will discuss below, $\tan\beta$ also becomes large in
these models.  The resulting formula for the one-loop corrected
Higgs mass in the limit $v_N \gg v$ and large $\tan\beta$, and
ignoring effects proportional to the relatively small stop mixing
parameter is given by
\begin{eqnarray}
m_{h_1}^2 &=&  M_Z^2\ - \frac{\lambda^4}{\kappa^2} v^2 + \delta
m_{h_1}^2 +
{\mathcal O}\left(\frac{v^4}{m_A^4}, \frac{1}{\tan\beta^2}
\right),
\label{404}\\
\delta m_{h_1}^2 &=& \frac{3m_t^4}{4\pi^2
v^2}\ln\left(\frac{\sqrt{m_{\tilde t_1}^2m_{\tilde
t_2}^2}}{m_t^2}\right) +{\mathcal O}\left(h_t^2 g^2,
h_t^2\lambda^2, \frac{A_t^2}{\sqrt{m_{\tilde t_1}^2 m_{\tilde
t_2}^2}}\right). \label{405}
\end{eqnarray}
Here $m_t$ is the running top quark mass at the top-quark mass
scale and
\begin{eqnarray}
m_A^2=\frac{2\mu(A_\lambda+ \frac{\kappa}{\lambda}\mu) +
\delta_A}{\sin 2\beta} \label{406}
\end{eqnarray}
is the would-be MSSM $CP$-odd Higgs boson mass, with
\begin{eqnarray}
\delta_A &\approx & \frac{3}{16 \pi^2} \frac{h_t^2A_t
\mu}{m_{\tilde t_1}^2-m_{\tilde t_2}^2} \left[m_{\tilde
t_1}^2\left(\ln\frac{m_{\tilde{t_1}}^2}{m_t^2}-1\right) -m_{\tilde
t_2}^2\left(\ln\frac{m_{\tilde{t_2}}^2}{m_t^2}-1\right)\right] \nonumber \\
&& +\frac{3}{16 \pi^2} \frac{h_b^2A_b \mu}{m_{\tilde
b_1}^2-m_{\tilde b_2}^2} \left[m_{\tilde
b_1}^2\left(\ln\frac{m_{\tilde{b_1}}^2}{m_t^2}-1\right) -m_{\tilde
b_2}^2\left(\ln\frac{m_{\tilde{b_2}}^2}{m_t^2}-1\right)\right]
\end{eqnarray}
being a one loop correction factor. Observe that we have omitted
the positive tree-level term proportional to $\lambda^2 \sin^2
2\beta$, which becomes unimportant for large values of
$\tan\beta$, and we have included the more important contribution
coming from the mixing with the singlet state, that in the limit
we are working becomes independent of the mass parameters of the
theory.  This occurs since the singlet $CP$-even state acquires a
mass about $4 \kappa^2 v_N^2$ and its mixing matrix element with
the lightest MSSM $CP$-even Higgs state is approximately equal to
$2 \lambda^2 v_N \; v$ in this limit. Note also that within this
approximation, the $\mathcal{O}(m_t^4)$ loop correction is
independent of the renormalization scale $\Lambda_{\overline{{\rm
MS}}}$, and is determined by the geometric average of the two stop
mass squares.

For the low-scale general gauge mediation, the RG evolution paths
of the stop soft masses $m_{\tilde Q_{3L}}^2$ and $m_{\tilde
t_R}^2$ are short. Given the effective SUSY breaking scales
$\Lambda_q \sim \Lambda_l \sim (10^5 -10^6)$ GeV, $m_{\tilde
t_1}^2$ and $m_{\tilde t_2}^2$, and hence $\delta m_{h_1}^2$ could
be large according to Eq.(\ref{405}). In such cases (see points
A2, A3, A4, A6 and A7, and also see points B4 and C4 in the
intermediate- and high-scale cases, respectively), the main
constraint on the model comes from the lightest chargino mass
$m_{\chi_1^c}$ which currently is bounded to be larger than 103.5
GeV~\cite{Chargino}. As emphasized above, to generate a large
chargino mass or effective $\mu$, we need to modify the relative
velocities of the RG evolutions of $m_{H_u}^2$ and $m_N^2$. With
$\eta$ shifted from $\sim 1$ to $\sim 2$, the RG evolution of
$m_{H_u}^2$ to a negative value is slowed down, but that of
$m_N^2$ is speeded up. A negative enough $m_N^2(\Lambda_{EW})$ and
hence a large enough $\mu$ are generated. For the intermediate- or
high-scale cases, because of the increased path length of its RG
evolution and the positivity of its beta function, $m_{\tilde
t_R}^2(\Lambda_{EW})$ becomes relatively small, leading to a small
$\delta m_{h_1}^2$. In these two cases (also see points A1, A5, A8
and A9 in the low-scale case), therefore, the main constraint on
the model comes from the lightest $CP$-even Higgs mass which
currently is bounded by 114.4 GeV~\cite{Higgs}. It is easy to see
according to Eq.(\ref{404}) that a small $\lambda$ and a large
$\kappa$ is helpful in obtaining a large $h_1$ mass. Let us stress
that in the numerical calculations of this paper, only the
dominant one-loop corrections to the Higgs effective potential
have been included. One should worry about the latent negative
effects on the Higgs masses from the higher-loop corrections which
may shift down the mass of the lightest $CP$-even Higgs by several
GeVs, similar to what happens in the MSSM (e.g.,
see~\cite{Espinosa:1991fc}). These negative effects may push the
lightest $CP$-even Higgs boson to values below the current
experimental bound.  This can be compensated, within our model, by
a slight shift in $\eta$ and a corresponding shift upwards of the
superparticle masses.



The general gauge mediation model discussed in the present work
favors heavy scalars and gauginos, as well as large values of
$\tan\beta$: $5 \sim 50$. The heaviness of the scalars and
gauginos of the theory is a reflection of the large values of the
effective SUSY breaking scales $\Lambda_{q,l}$ necessary to
fulfill the Higgs and/or chargino mass constraints. The preference
for large values of $\tan\beta$ can be easily understood by
analyzing the minimization conditions. First of all, $B_\mu$ is
relatively small because the boundary value of the soft parameter
$A_\lambda$ at the messenger scale is highly suppressed in our
model. Then, since the term
$m_{H_d}^2(\Lambda_{EW})+m_{H_u}^2(\Lambda_{EW})$ in
Eq.(\ref{315}) is typically larger than the other terms in
Eq.(\ref{315}) (see Table(\ref{table1}), (\ref{table3}), and
(\ref{table5})), only a relatively large $\tan\beta$ can suppress
the RHS of Eq.(\ref{315}) to make it match with a small $B_\mu$.
The precise value of $\tan\beta$ depends on the messenger scale. A
higher messenger scale $\Lambda_M$ generally leads to a more
negative $m_N^2(\Lambda_{EW})$ because of the extended RG
evolution path (actually, the enlarged beta function of $m_N^2$
due to a larger $\eta$ required by phenomenology also has a
contribution.) or a larger $\kappa v_N$ according to
Eq.(\ref{316}). According to Eq.(\ref{301}) and Eq.(\ref{315}),
this indicates a larger $B_\mu$ or equivalently, a smaller
$\tan\beta$.  Therefore, for fixed $\lambda$ and $\kappa$,
$\tan\beta$  becomes smaller as $\Lambda_M$ increases. On the
other hand, for fixed $\Lambda_M$, a larger $\tan\beta$ often
implies a larger $\lambda$ or a smaller $\kappa$. For fixed
$\kappa$, a larger $\lambda$ implies a larger beta function for
$m_N^2$ or a more negative $m_N^2(\Lambda_{EW})$, so a smaller
$\tan\beta$ can be explained according to the same argument as
that in case. For fixed $\lambda$, a smaller $\kappa$ implies a
larger $v_N$ or $B_\mu$ according to Eq.(\ref{316}) and
Eq.(\ref{301}). This then leads to a smaller $\tan\beta$ again
according to Eq(\ref{315}).

A large $\tan\beta$ is welcomed in phenomenology, due to its role
in explaining the mass hierarchy of top and bottom quarks or
realizing the unification of their Yukawa couplings (e.g.,
see~\cite{Hall:1993gn}). In our model, relatively large values of
$\tan\beta$ bring us more than that,
since it helps in avoiding an
unacceptably light chargino: in the mass formula of the lightest
chargino Eq.(\ref{403}), the corrections at the order
$\mathcal{O}(\frac{\mu}{M_2})$ contain a negative contribution
\begin{eqnarray}
\mathcal{O}\left(\frac{\mu}{M_2}\right) \simeq  - \frac{2\mu
}{|M_2|}m_W^2\sin 2\beta, \label{407}
\end{eqnarray}
which is suppressed by a large $\tan\beta$. Moreover, a relatively
large $\tan\beta$ plays a crucial role in the solution of the
light $U(1)_R$ peudoscalar problem.

As first pointed out in~\cite{Dine:1993yw}, small
$|A_\lambda(\Lambda_{EW})|$ and $|A_\kappa(\Lambda_{EW})|$
(compared to $\Lambda_{EW}$) induce the presence of a light
pseudoscalar. In this limit, the mass of the lightest $CP$-odd
Higgs boson is approximately given by (e.g.,
see~\cite{Dobrescu:2000yn}):
\begin{equation}
m_{a_1}^2  = 3v_N \left( \frac{3 \lambda A_\lambda \cos^2 \theta_A
}{2 \sin 2 \beta}
 + \kappa  A_\kappa \sin^2 \theta_A \right) + \mathcal{O}\left(\frac{A_\lambda}{v},
\frac{A_\kappa}{v}\right), \label{113}
\end{equation}
where
\begin{equation}
a_1 =  \cos \theta_A \, A_{MSSM} + \sin \theta_A \, A_N
\label{114}
\end{equation}
with $A_{MSSM}$ and $A_N$ being the doublet and singlet $CP$-odd
gauge eigenstates, respectively, and $0 \leq \theta_A \leq
\frac{\pi}{2}$ being their mixing angle.  Depending on its
composition, a light pseudoscalar may be in conflict with the
strong LEP bounds. As extensively discussed in the literature,
this light pseudoscalar should be understood as the
Nambu-Goldstone boson of the global $U(1)_R$ symmetry, since
$A_\lambda(\Lambda_{EW})$ and $A_\kappa(\Lambda_{EW})$ represent
the only two terms explicitly violating this symmetry. However,
from Table (\ref{table1})-(\ref{table8}), it is easy to see that
there is no such a problem in our model: for the intermediate- and
high-scale gauge-mediations, $|A_\lambda(\Lambda_{EW})|$ is
typically large, compared to $\Lambda_{EW}$; for the low-scale
case, even though $|A_\lambda(\Lambda_{EW})|$ and
$|A_\kappa(\Lambda_{EW})|$ are small (except point A8), the light
pseudoscalar is extremely singlet-like (see Table(\ref{table7})),
escaping the experimental constraints successfully.

These features are due to $\eta$ and the relatively large
$\tan\beta$ again. Consider the strongly coupled RG evolutions of
$A_t$, $A_b$ and $A_\lambda$ (see RG equations (\ref{A110}),
(\ref{A111}) and (\ref{A113})).
At the messenger scale we have $A_t(\Lambda_M)\sim
A_b(\Lambda_M) \sim A_\lambda(\Lambda_M) \sim 0$ in our model.
A larger $\eta$ implies more negative contributions to the beta
functions according to the EW gaugino soft masses, and less
negative contributions according to the gluino soft mass. Since
the latter is absent in the beta function of $A_\lambda$, but
contributing to those of $A_t$ and $A_b$, a large $\eta$
necessarily leads to a larger $A_\lambda(\Lambda_{EW})$, as long
as the evolution pathes are long enough. This explains the
relatively large $A_\lambda(\Lambda_{EW})$ and large $U(1)_R$
peudoscalar masses in the contexts of the intermediate- and
high-scale general gauge mediations. Unlike these two cases, the
$U(1)_R$ peudoscalar is still light in the low-scale case (except
point A8) due to the short RG evolution path for $A_\lambda$. A
relatively large $\tan\beta$ plays a crucial role in avoiding the
experimental bound here. As shown in~\cite{Dobrescu:2000yn}, the
mixing angle $\theta_A$ of the $U(1)_R$ pseudoscalar $a_1$
satisfies
\begin{eqnarray}
\tan \theta_A = \frac{v_N}{v\sin 2\beta} +
\mathcal{O}(\frac{A_\lambda}{v},\frac{A_\kappa}{v})
\end{eqnarray}
under the limit of small $A_\lambda(\Lambda_{EW})$ and
$A_\kappa(\Lambda_{EW})$. Obviously, a relatively large $\tan
\beta$ implies
\begin{eqnarray}
\theta_A \approx \frac{\pi}{2}
\end{eqnarray}
and hence an extremely singlet-like $U(1)_R$ pseudoscalar $a_1$
(see Table(\ref{table7})). This is also true for the few examples
in the intermediate-scale general gauge mediation (points B1, B2
and B3 in Table (\ref{table7})). The light $U(1)_R$ pseudoscalar
problem, therefore, is no longer a problem in our model.


\begin{figure}
\caption{Boundary between the perturbative and the
non-perturbative regions on the
$\lambda(\Lambda_{EW})-\kappa(\Lambda_{EW})$ plane. In the
perturbative region (blank one), $\lambda$ and $\kappa$ keep
perturbative up to the GUT scale. The stars on the plane denote
the sample points we are studying. The boundary has a weak
dependence on Yukawa couplings and the messenger scale. Here we
set $h_t({\Lambda_{EW}})=0.95$, $h_b(\Lambda_{EW})=0.5$ and
$\Lambda_M=10^{11}$ GeV.} \label{fig1}
\begin{center}
\includegraphics[height=4.5 in]{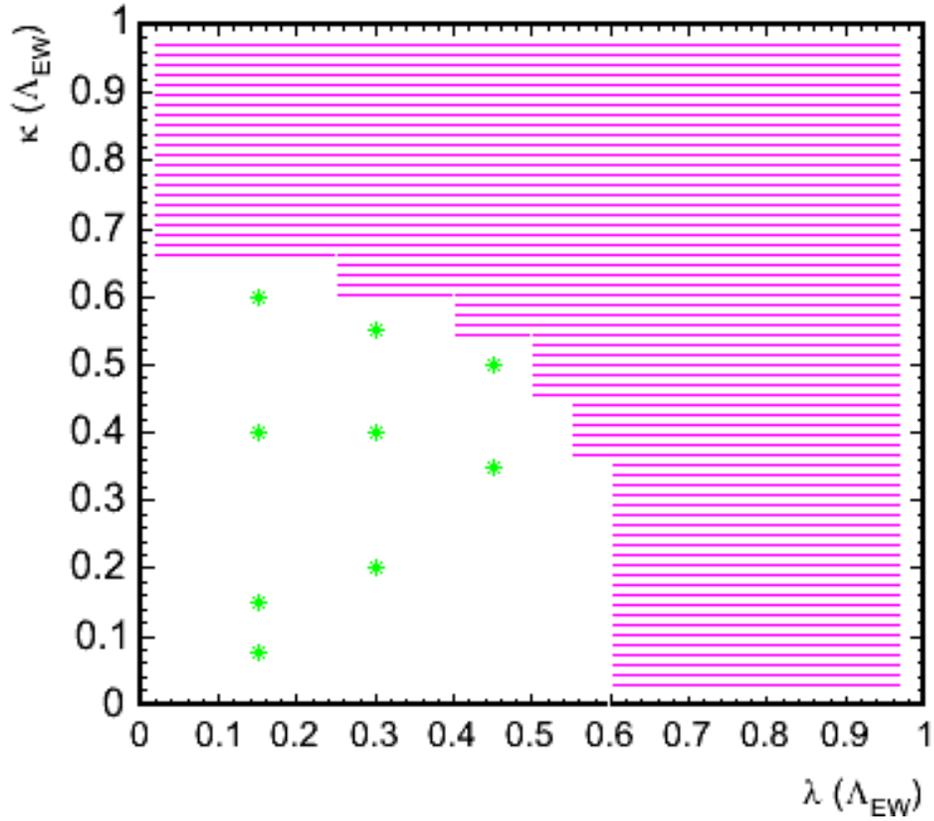}
\end{center}
\end{figure}

To end this section, let us take a look at the possible range of
$\lambda$ and $\kappa$ at the EW scale in the NMSSM. The most
serious constraint is from the requirement of $\lambda$ and
$\kappa$ to be perturbative up to the GUT scale. For the case
where the gauge couplings are the only possible tree-level
interactions between the observable and messenger sectors, the
boundary between the perturbative and non-perturbative regions has
been drawn in Fig.(\ref{fig1}), with $h_t({\Lambda_{EW}})=0.95$,
$h_b(\Lambda_{EW})=0.5$ and $\Lambda_M=10^{11}$ GeV. From the
figure, it is easy to see that both large $\lambda(\Lambda_{EW})$
and large $\kappa(\Lambda_{EW})$ regions have been excluded, and
the only allowed region is located in the lower-left corner of the
$\lambda(\Lambda_{EW})-\kappa(\Lambda_{EW})$ plane. The boundary
in the figure depends on the Yukawa couplings as well as the
messenger scale. But this dependence is very weak: there is only a
mild shift as these parameters vary in the region in which we are
interested. The stars on the
$\lambda(\Lambda_{EW})-\kappa(\Lambda_{EW})$ plane denote the
sample points we are studying in this paper. It is easy to see
that these points cover almost the whole perturbative region on
the $\lambda(\Lambda_{EW})-\kappa(\Lambda_{EW})$ plane. In
particular, all of them lead to reasonable particle mass spectra
which satisfy the current experimental bounds\footnote{Actually,
if the requirements of perturbativity (up to $\Lambda_{GUT}$) for
the couplings are given up, the nice features of these examples
could be extended into the non-perturbative region on the
$\lambda(\Lambda_{EW})-\kappa(\Lambda_{EW})$ plane as long as
these couplingss stay perturbative at the messenger scale. }.
Therefore, the $\mu/B_{\mu}$ problem is solved in the context of
the general gauge mediated SUSY breaking model analyzed in this
work.

\section{Collider Signals}

Although a detailed analysis of the collider signatures of these
models is beyond the scope of this article, we would like to
stress some relevant properties of these models and their
associated phenomenology.

For the low scale gauge mediation, all colored particles are very
heavy and therefore very difficult to detect at hadron colliders.
One promising way to test these models is by analyzing the
production and two-body decay of the next-to-lightest
superparticle (NLSP) to gravitino ($\tilde G^\alpha$) which is
described by
\begin{eqnarray}
\mathcal{L} \sim \frac{1}{F}\partial_\mu G^\alpha j^\mu_\alpha +
h.c. , \label{421}
\end{eqnarray}
Here $\sqrt{F}$ is the SUSY breaking scale, and $j^\mu_\alpha$ is
the supercurrent. In our model, the NLSP generally is the lightest
neutralino which is typically Higgsino-like, and in most cases
whose mixing with singlino is suppressed (see Table(\ref{table8})
for the case with low-scale gauge mediation). So the most
important experimental signature would be the di-$Z$ and di-$h_1$
productions (if allowed by phase space)
\begin{eqnarray}
 \chi_1^0 &\rightarrow & Z \tilde G : \ \ \ \ Z Z + X + \not\! E
\nonumber \\  \chi_1^0 &\rightarrow & h_1 \tilde G : \ \ \ \
h_1h_1 + X + \not\! E \label{115}
\end{eqnarray}
here $X$ is any collection of leptons and jets, and $\not\! E$
denotes the missing energy. Explicitly, under the Higgsino-like
limit (with the mixing with the singlino suppressed), the decay
rates to $Z$-boson and $h_1$ are given by~\cite{Ambrosanio:1996jn}
\begin{eqnarray}
\Gamma( \chi^0_1 \rightarrow Z \tilde G) &\approx&
\frac{1}{2}|c_{\tilde H_d} \cos\beta + c_{\tilde H_u} \sin \beta
|^2\frac{m_{\chi^0_1}^5}{16\pi F^2}
\left(1-\frac{m_Z^2}{m_{\chi^0_1}^2} \right)^4, \nonumber \\
\Gamma( \chi^0_1 \rightarrow h_1 \tilde G) &\approx&
\frac{1}{2}|c_{\tilde H_d} \sin \alpha - c_{\tilde H_u}\cos
\alpha|^2\frac{m_{\chi^0_1}^5}{16\pi F^2}
\left(1-\frac{m_{h_1}^2}{m_{\chi^0_1}^2} \right)^4 .
\end{eqnarray}
Here $c_{\tilde H_d}$ and $c_{\tilde H_u}$ are composition
coefficients of $\chi_1^0$, and $\alpha$ is the $h_1-h_2$ Higgs
mixing angle.
If the singlino component of $\chi_1^0$ is not
small, e.g., at A1 point, the di-$a_1$ decay can also provide
useful collider signals.
Note that in our
model independently of that $\langle S_q\rangle \sim \langle
S_l\rangle \sim \Lambda_M$ or $\langle S_q\rangle \sim \eta
\langle S_l\rangle \sim \Lambda_M^2/\langle S_l\rangle$ is
assumed, we typically have $\sqrt{F}\sim \sqrt{F_qF_l}$ between a
few $10^5$ GeV and a few $10^6$ GeV, which implies non-prompt
di-boson decays~\cite{Culbertson:2000am}. This is important since
the background for any of the final state signatures can be
greatly reduced (due to the displaced vertices and distinguished
angular distribution of the displaced jets from $Z$ or $h_1$
decays) if the $\chi^0_L$ decay is non-prompt but contained in the
tracking region.

It is also important to stress that in the low scale gauge
mediated scenario, the Higgs decays may be affected by the
presence of the light pseudoscalars, $a_1$. Although the lightest
pseudoscalar is mostly a singlet state (see Table~\ref{table7}),
it will decay into bottom quark and $\tau$ pairs through its
mixing with the pseudoscalar component of the Higgs doublet $H_d$.
Therefore, the Higgs decay into two $a_1$ states will induce
decays into either four bottom quarks, two bottom quark and two
$\tau$'s, or four $\tau$'s final states. The final signatures of
the di-$h_1$ channel in \ref{115}, necessarily, will also be
affected. In all the scenarios we presented, the lightest CP-even
Higgs is sufficiently heavy as to evade the stringent LEP
constraints on a light CP-even Higgs decaying into four bottom
quark final states~\cite{htoaaLEP}. The presence of these new
decay channels will demand new strategies for the search for
CP-even Higgs bosons at the Tevatron and the LHC, as has been
recently analyzed in Refs.~\cite{htoaa}.

The gravitino collider signals are seriously suppressed for
intermediate- and high-scale gauge mediations, since the
neutralino lifetime will be enhanced by the factor $F^{2}$ and
therefore it will decay beyond the detector. Moreover, whenever
light, the charged and neutral Higgsinos would be approximately
degenerate in mass and therefore difficult to detect by direct
production at hadron colliders. However, colored particles become
lighter and therefore they provide the most important search
channels at the LHC. In the high-scale case, the gluino mass
$m_{\tilde g}$ is typically around 1.5 TeV or even smaller,
implying an abundant production of gluinos at LHC, according to
the gluino ($\tilde g$) pair production
\begin{eqnarray}
pp \rightarrow \tilde g \tilde g.
\end{eqnarray}
Meanwhile, given that the lightest stop $\tilde t_1$ is mainly
right-handed and much lighter than gluino in this case, one could
expect to see the signatures at LHC according to the decaying
channels
\begin{eqnarray}
\tilde g & \rightarrow & t\tilde t_1 \ \ \rightarrow \ \  tt
\chi_1^0, \nonumber \\
\tilde g & \rightarrow & t\tilde t_1 \ \ \rightarrow \ \  tb
\chi_1^c.
\end{eqnarray}
Therefore, the final state will be given by four top quarks or two
top and two bottom quarks with large missing energy. An analysis
of similar gluino decay channels at the LHC has been performed in
Ref.~\cite{Kitano}. Even though we typically have $m_{\chi_1^0}<
m_{\tilde t_1}$ in the high-scale scenario, C9 point is an
exception, where $\tilde t_1$ is lighter than $\chi_1^0$ and
$\tilde \tau_1$. The light stop $\tilde t_1$ is long-lived because
its two-body decay to gravitino
\begin{eqnarray}
\tilde t_1 & \rightarrow & t\tilde G,
\end{eqnarray}
even if kinematically allowed, is also suppressed by a $F^{-2}$
factor. In such a case, the stop may have interesting implications
on both cosmology and collider signatures. For more details,
readers may refer to~\cite{Santoso:2007uw}.

As for the intermediate-scale scenario, even though an abundant
production of gluinos at LHC is also expected for many cases, the
mass of the lightest stop is typically larger than that of
gluinos. Whenever the gluino mass is within kinematic reach of
the LHC,
they  will decay only through off-shell squarks
\begin{eqnarray}
\tilde g \rightarrow qq' \chi_i^0, \ \ \ \ \tilde g \rightarrow
qq'\chi^c_i.
\end{eqnarray}
Since the neutralinos and charginos appearing in the intermediate
states have multiple decay modes, there will be many competing
gluino decay chains whose branching ratios are quite sensitive to
the parameters of this model. Interested readers may refer
to~\cite{Toharia:2005gm} and its references.

\section{Discussions and Conclusions}

The general gauge mediation provides a simple way to solve the
$\mu/B_\mu$ problem in the NMSSM. In this context, reasonable
values for $\mu/B_\mu$ can be generated by properly modifying the
RG evolutions of $m_{H_u}^2$ and $m_N^2$ by a choice of $\eta$
window. The EW scale is then stabilized, and phenomenologically
interesting spectra of particles and superparticles are also
achieved. These features apply to most of the perturbative (up to
the GUT scale) $\lambda-\kappa$ parameter region in the NMSSM and
to all phenomenologically interesting messenger scales. In
addition, there is no light $U(1)_R$ pseudoscalar problem in our
model. For the intermediate- and high-scale gauge-mediations, due
to a relatively heavy spectrum of gauginos, large
$|A_\lambda(\Lambda_{EW})|$ or $|A_\kappa(\Lambda_{EW})|$,
comparable with $\Lambda_{EW}$ are typical, so the lightest Higgs
pseudoscalar is not too light. For the low-scale case, even though
$|A_\lambda(\Lambda_{EW})|$ and $|A_\kappa(\Lambda_{EW})|$ are not
always large, the lightest Higgs pseudoscalar is extremely
singlet-like due to a relatively large $\tan\beta$ favored by our
model, escaping the experimental constraints on a light Higgs
boson.

It is worth emphasizing that the
introduction of the parameter $\eta$ does not affect the
successful prediction of the gauge coupling unification at the GUT
scale. Recall the threshhold corrections to the gauge coupling
unification due to the little hierarchy between the EW scale and
the soft SUSY breaking scale, where
$\frac{\Lambda_{soft}}{\Lambda_{EW}}\sim 10$ and many charged
particle species are involved. In the general gauge mediation
scenario described in this article,
the correction to the prediction of
$\alpha_3(M_Z)$ induced by the messenger threshold corrections
may be estimated by
\begin{eqnarray}
\Delta \alpha_3(M_Z) \simeq \frac{9}{14 \pi} \; \alpha_3(M_Z)^2 \;
\ln\left(\frac{\langle S_q \rangle}{\langle S_l \rangle} \right)
\label{eq:Thalpha3}
\end{eqnarray}
On the other hand, the introduction of $\eta$ also modifies the
sparticle threshold corrections, which are approximately given by~\cite{CPW}
\begin{eqnarray}
\Delta \alpha_3(M_Z) & \simeq & -\frac{19}{28 \pi} \;
\alpha_3(M_Z)^2 \; \ln\left( \frac{|\mu|}{M_Z}
\left(\frac{M_2}{M_3}\right)^{3/2} \right)
\nonumber\\
& = & - \frac{19}{28 \pi} \alpha_3(M_Z)^2 \ln\left(
\frac{|\mu|}{M_Z} \left(\frac{\eta \;
\alpha_2}{\alpha_3}\right)^{3/2} \right) \label{eq:CPW}
\end{eqnarray}

One could compute the difference in the prediction of
$\alpha_3(M_Z)$ with respect to the case $\eta = 1$. Let us
consider two cases. In the first one, the ratio of effective SUSY
breaking scales $\Lambda_l/\Lambda_q \sim \langle F_l \rangle/
\langle F_q \rangle \simeq \eta$, and therefore $\langle S_q
\rangle/\langle S_l \rangle \sim 1$.  In such a case,
\begin{equation}
\Delta_{\eta} \alpha_3(M_Z) \simeq -
\frac{57}{56 \pi}
\; \alpha_3^2(M_Z) \; \ln \eta .
\end{equation}
Alternatively, one can consider $\langle F_q \rangle \sim \langle
F_l \rangle$ and therefore $\langle S_q \rangle /\langle S_l
\rangle \simeq \eta$.  In this case,
\begin{equation}
\Delta_{\eta} \alpha_3(M_Z) \simeq -
\frac{21}{56 \pi}
\; \alpha_3^2(M_Z) \;
\ln \eta .
\end{equation}
In both cases, the total correction is negative, leading, for
$\eta \simeq 2$--$6$ to a somewhat better
agreement between the predicted and measured values of $\alpha_3(M_Z)$
than in the $\eta = 1$ case~\footnote{Successful unification in
the $\eta = 1$ case requires
the threshold scale $|\mu| (\alpha_2/\alpha_3)^{3/2} \simeq 1 TeV$.}.

One interesting feature on this model is the arising of one
physical $CP$-phase according to the gaugino soft masses. In the
NMSSM with general gauge mediation, there are four independent
complex parameters: $\lambda$, $\kappa$, and two of the soft
gaugino masses $M_1$, $M_2$ and $M_3$. Among them, the phase of
$\lambda$ is not physical and can be resolved by the CKM matrix.
In addition, $\kappa$ and gaugino soft mass are not invariant
under the Peccei-Quinn symmetry and $U(1)_R$ symmetry,
respectively. The phase of $\kappa$ and one phase in the gaugino
mass sector hence can be rotated away. So there is one physical
phase left in the soft mass sector of gauginos. On the other hand,
it is well-known that the CKM phase is not enough and extra
$CP$-violating sources are required to explain the origin of the
baryon asymmetry in the Universe today. The physical $CP$ phase
appearing in our model may provide a nice chance to understand
this cosmic mystery. For example, in the EW baryogenesis mechanism
(see~\cite{Riotto:1999yt} for a review or \cite{Carena:1997gx} for
its realization in different supersymmetric models), such a phase
may induce a net amount of left-chiral weak fermions during the EW
phase transition, which is then switched to the baryon asymmetry
in the Universe according to the EW sphaleron effect. But, the
same as the $CP$ phases appearing in any other supersymmetric
models, the physical $CP$-phase in our model also needs to satisfy
the EDM bounds of electron, neutron and mercury atoms. Since the
masses of the first two family squarks in our model are typically
heavier than $2-3$ TeV, it might be viable to suppress its
one-loop contributions to the EDMs according to the heavy squark
mechanism~\cite{Kizukuri:1992nj}. In addition, it is claimed
recently~\cite{Suematsu:2007iv} that in a context similar to ours,
a large $CP$-phase of order $\mathcal{O}(1)$ can be consistent
with all EDM bounds according to some cancellation effects, with
no necessity to require the squarks of the first two families to
be heavy.

It is interesting to ask why the situation is so different
between the class of models~\cite{Giudice:1997ni,Delgado:2007rz}
and our model, since both of them have a total of four free input
parameters,  with one messenger coupling in the former case
replaced by the parameter $\eta$ in our model. To great extent
this is due to the different ways in which  the
negative soft mass square $m_N^2(\Lambda_{EW})$ is generated. In the former
case, the authors try to generate a negative $m_N^2(\Lambda_{EW})$
directly according to the boundary conditions at the messenger
scale. They let the singlet $\mathbf{N}$ directly couple to the
messengers. Then, the contribution of this coupling
to $m_N^2(\Lambda_M)$ at two-loop level are negative.
But this coupling has similar negative contributions to
$m_{H_u}^2(\Lambda_M)$
making $m_{H_u}^2$ get enough negative values quickly to induce EW
symmetry breaking. This in turn refrains $m_N^2$ from getting a
too negative value at the EW scale according to the RG evolution.
In our model, we try to generate a negative $m_N^2(\Lambda_{EW})$
by modifying the related RG evolutions. The introduced parameter
$\eta$ have opposite effects on the beta functions $m_{H_u}^2$ and
$m_N^2$, so the evolution of $m_N^2$ to a negative value is
accelerated while that of $m_{H_u}^2$ is slowed down. This allows
$m_N^2$ have enough time to obtain a very negative value before
$m_{H_u}^2$ induces the EW symmetry breaking. In addition, unlike
the former case, the trilinear soft parameters $A_\lambda$ and
$A_\kappa$ are highly suppressed at the messenger scale, which
leads to a relatively large $\tan\beta$ at the EW scale. This
relatively large $\tan\beta$ not only helps lift the mass of the
lightest chargino, but more importantly, help solve the light
$U(1)_R$ pseudoscalar problem by suppressing its mixing with the
SM-like $CP$-odd Higgs components.

The general gauge mediation is natural because of its simplicity
and universality. It is very simple, only requiring minimal
messenger spectrum in the messenger sector and with no additional
symmetry or new dimensional parameters introduced. Most
importantly, it can naturally arise from a general hidden sector,
as pointed out in subsection~\ref{the model}. Since the
construction of this model is independent of the visible sector,
its idea can also be extended to many other contexts without much
difficulty, e.g., the nMSSM and the UMSSM, or even help the class
of models in~\cite{Giudice:1997ni,Delgado:2007rz} obtain more
reasonable physical results. Due to the similar structures of the
related beta functions, we believe that similar effects could be
seen in these extensions. We will leave these interesting issues
to future exploration.


\section*{Acknowledgments}

We would like to thank Andr\'{e} de Gouv\^{e}a, John Gunion, Paul
Langacker, Arjun Menon and David Morrissey for useful discussions
and comments. Work at ANL is supported in part by the U.S.
Department of Energy (DOE), Div. of HEP, Contract
DE-AC02-06CH11357. Work at EFI is supported in part by the U.S.
DOE through Grant No. DEFG02- 90ER40560. T.L. is also supported by
Fermi-McCormick Fellowship.

\newpage

\appendix

\renewcommand{\thesection}{Appendix \Alph{section}}

\setcounter{equation}{0}
\section{The RG Equations in the NMSSM}

\renewcommand{\theequation}{\Alph{section}.\arabic{equation}}

All of the one-loop RG equations in the NMSSM (e.g.,
see~\cite{King:1995vk}) are listed in this section. Considering
that the beta functions of RGEs in a general background also
depend on the couplings between the observable sector, and the
messenger and hidden sectors, here we assume that at tree-level
there is no couplings between the observable and the hidden
sector, and the gauge couplings are the only possible interactions
between the observable and the messenger sector. In addition, in
the numerical work of this paper, we neglect all threshhold
corrections to the RG evolutions caused by the little hierarchy
between the EW scale ($\sim 100$ GeV) and the soft SUSY breaking
scale ($\sim 1000$
GeV). \\

{\bf I. The sector of superpotential couplings}

\begin{eqnarray}
16\pi^2\frac{d}{dt} g_Y     & = & 11g_Y^3 , \\
16\pi^2\frac{d}{dt} g_2     & = & g_2^3 , \\
16\pi^2\frac{d}{dt} g_3     & = & -3g_3^3  ,\\
16\pi^2\frac{d}{dt} h_t     & = & (6 h_t^2 + h_b^2 + \lambda^2
                   - \frac{13}{9} g_Y^2
                   - 3 g_2^2 - \frac{16}{3} g_3^2) h_t ,\\
16\pi^2\frac{d}{dt} h_b     & = & (6 h_b^2 + h_t^2 + h_\tau^2 +
\lambda^2
                   - \frac{7}{9} g_Y^2 - 3 g_2^2
                   - \frac{16}{3} g_3^2) h_b ,\\
16\pi^2\frac{d}{dt} h_\tau  & = & (4 h_\tau^2 + 3 h_b^2 +
\lambda^2
                   - 3 g_Y^2 - 3 g_2^2) h_\tau ,\\
16\pi^2\frac{d}{dt} \lambda & = & (4 \lambda^2 + 2 k^2 + 3 h_t^2
                   + 3 h_b^2 + h_\tau^2
                   - g_Y^2 - 3 g_2^2) \lambda ,\\
16\pi^2\frac{d}{dt} k       & = & 6 (\lambda^2 + k^2) k .
\end{eqnarray}
In the above equations $g_Y=e /\cos \theta_{EW}$ is the $U(1)_{Y}$
gauge coupling. In the GUT framework, it is generally normalized
to be $g_1 \equiv \sqrt {\frac{5}{3}} g_Y$ and $\alpha_1 \equiv
\frac{5}{3} \alpha_Y$. When $t>{\rm
ln}(\frac{\Lambda_M}{\Lambda_{EW}})$, the RGEs of $g_Y$, $g_2$ and
$g_3$ are modified to
\begin{eqnarray}
16\pi^2\frac{d}{dt} g_Y     & = & (11+\frac{5n}{3})g_Y^3 , \\
16\pi^2\frac{d}{dt} g_2     & = & (1+n)g_2^3 , \\
16\pi^2\frac{d}{dt} g_3     & = & (-3+n)g_3^3
\end{eqnarray}
with n being the number of messenger pairs $(3+2)+(\bar 3+\bar 2)$.\\

{\bf II. The sector of soft A-term couplings}

\begin{eqnarray}
16\pi^2\frac{d}{dt} A_{u_a}  & = & 6 h_t^2 (1+\delta_{a 3}) A_t
                 + 2 h_b^2 \delta_{a 3} A_b
                  + 2 \lambda^2 A_\lambda \nonumber \\
            && -  4 (\frac{13}{18} g_Y^2 M_1 + \frac{3}{2} g_2^2 M_2
                       + \frac{8}{3} g_3^2 M_3) , \label{A110}\\
16\pi^2\frac{d}{dt} A_{d_a} & = & 6 h_b^2 (1+\delta_{a 3}) A_b
                  + 2 h_t^2\delta_{a 3} A_t
                  + 2 h_\tau^2 \delta_{a 3} A_\tau
                  + 2 \lambda^2 A_\lambda \nonumber \\
            && -  4 (\frac{7}{18} g_Y^2 M_1 + \frac{3}{2} g_2^2 M_2
                       + \frac{8}{3} g_3^2 M_3) ,\label{A111}\\
16\pi^2\frac{d}{dt} A_{e_a} & = & 2 h_\tau^2 (1 + 3 \delta_{a 3})
A_\tau
                +  6 h_b^2 A_b + 2 \lambda^2 A_\lambda \nonumber \\
            && -  6 (g_Y^2 M_1 + g_2^2 M_2) ,\label{A112} \\
16\pi^2\frac{d}{dt} A_\lambda & = & 8 \lambda^2 A_\lambda - 4 k^2
A_k
                    + 6 h_t^2 A_t
                    + 6 h_b^2 A_b + 2 h_\tau^2 A_\tau \nonumber \\
            && -  2 (g_Y^2 M_1 + 3 g_2^2 M_2) ,\label{A113}\\
16\pi^2\frac{d}{dt} A_k     & = & 12 (k^2 A_k - \lambda^2
A_\lambda). \label{A114}
\end{eqnarray}
Here $A_{i}$ are the soft SUSY-breaking A-term couplings. $M_i$
($i$=1,2,3) are the soft SUSY-breaking gaugino masses which evolve
as
\begin{eqnarray}
M_1(t) &=&  \frac{{g_Y(t)}^2}{16\pi^2} (\Lambda_l+\frac{2}{3}\Lambda_q)  \label{A115} \\
M_2(t) &=&  \frac{{g_2(t)}^2}{16\pi^2}\Lambda_l  \label{A116} \\
M_3(t) &=&  \frac{{g_3(t)}^2}{16\pi^2}\Lambda_q    \label{A117}
\end{eqnarray}
at one-loop level in our model.\\

{\bf III. The sector of soft SUSY-breaking masses}

\begin{eqnarray}
16\pi^2\frac{d}{dt} m_{\tilde{Q}_a}^2 & = & 2 \delta_{a 3} h_t^2
                    (m_{\tilde{Q}_3}^2 + m_{H_u}^2 + m_{\tilde{t}}^2 + A_t^2)
                  + 2 \delta_{a 3} h_b^2
                    (m_{\tilde{Q}_3}^2 + m_{H_d}^2 + m_{\tilde{b}}^2 +
                    A_b^2)
                    \nonumber \\
              & &-  8 (\frac{1}{36} g_Y^2 M_1^2 + \frac{3}{4} g_2^2 M_2^2
                      +\frac{4}{3} g_3^2 M_3^2)
               ,\\
16\pi^2\frac{d}{dt} m_{\tilde{u}_a}^2 & = & 4 \delta_{a 3} h_t^2
                    (m_{\tilde{Q}_3}^2 + m_{H_u}^2 + m_{\tilde{t}}^2 +
                    A_t^2)
                  -  8 (\frac{4}{9} g_Y^2 M_1^2 + \frac{4}{3} g_3^2 M_3^2)
             , \\
16\pi^2\frac{d}{dt} m_{\tilde{d}_a}^2 & = & 4 \delta_{a 3} h_b^2
                    (m_{\tilde{Q}_3}^2 + m_{H_d}^2
                    + m_{\tilde{b}}^2 + A_b^2)  -  8 (\frac{1}{9} g_Y^2 M_1^2 + \frac{4}{3} g_3^2 M_3^2)
                   ,\\
16\pi^2\frac{d}{dt} m_{\tilde{L}_a}^2 & = & 2 \delta_{a 3}
h_\tau^2
                    (m_{\tilde{L}_3}^2 + m_{H_d}^2
                    + m_{\tilde{\tau}}^2 + A_\tau^2) -  8 (\frac{1}{4} g_Y^2 M_1^2 + \frac{3}{4} g_2^2 M_2^2)
                  ,\\
16\pi^2\frac{d}{dt} m_{\tilde{e}_a}^2 & = & 4 \delta_{a 3}
h_\tau^2
                    (m_{\tilde{L}_3}^2 + m_{H_d}^2
                    + m_{\tilde{\tau}}^2 + A_\tau^2)  -  8 g_Y^2 M_1^2 ,\\
16\pi^2\frac{d}{dt} m_{H_d}^2 & = & 6 h_b^2
                    (m_{\tilde{Q}_3}^2 + m_{H_d}^2 + m_{\tilde{b}}^2 + A_b^2)
                  + 2 h_\tau^2
                    (m_{\tilde{L}_3}^2 + m_{H_d}^2
                    + m_{\tilde{\tau}}^2 + A_\tau^2) \nonumber \\
              && +  2 \lambda^2
                    (m_{H_d}^2 + m_{H_u}^2 + m_N^2 + A_\lambda^2)
                   - 8(\frac{1}{4} g_Y^2 M_1^2 + \frac{3}{4} g_2^2 M_2^2)
                  ,\label{A103}\\
16\pi^2\frac{d}{dt} m_{H_u}^2 & = & 6 h_t^2
                   (m_{\tilde{Q}_3}^2 + m_{H_u}^2 + m_{\tilde{t}}^2 + A_t^2)
                  + 2 \lambda^2
                   (m_{H_d}^2 + m_{H_u}^2 + m_N^2 + A_\lambda^2)
                    \nonumber \\
              && -  8 (\frac{1}{4} g_Y^2 M_1^2 + \frac{3}{4} g_2^2 M_2^2)
            ,   \label{A101}\\
16\pi^2\frac{d}{dt} m_{N}^2   & = & 4 \lambda^2
                    (m_{H_d}^2 + m_{H_u}^2 + m_N^2 + A_\lambda^2)
                   + 4 k^2 (3 m_N^2 + A_k^2). \label{A102}
\end{eqnarray}
Here all soft SUSY-breaking masses are taken to be diagonal. Note,
in all of the three sectors, only the effect of the third
generation Yukawa couplings, $i.e.$, $h_{t}$, $h_{b}$ and
$h_{\tau}$ are considered.

\section{Numerical Results}

In this section, we list the numerical results in the cases of
low- (Table(\ref{table1})-(\ref{table2})), intermediate-
(Table(\ref{table3})-(\ref{table4})), and high-scale
(Table(\ref{table5})-(\ref{table6})) general gauge mediations. The
composition of the light $U(1)_R$ pseudoscalar is given in
Table(\ref{table7}), and the composition of the lightest
neutralino or the NLSP in the low-scale case is given in
Table(\ref{table8}).

\begin{table}
\caption{Parameters of the low-scale general gauge mediation. }
\label{table1}
\bigskip
\begin{center}
\noindent{
\begin{tabular}{|c|c|c|c|c|} \hline
 & \multicolumn{4}{|c|}{Input Parameters} \\ \hline
{\rm Pts} & $\lambda(\Lambda_{EW})$ & $\kappa(\Lambda_{EW})$ & $\Lambda_M$ (GeV) & $\eta$   \\
 \hline

A1 & 0.15 & 0.075 & $2.50\times10^5$ & 2.1160   \\
 \hline

A2 & 0.15 & 0.15 & $5.00\times10^5$ & 2.2708  \\
 \hline

A3 & 0.15 & 0.40 & $5.00\times10^6$ & 2.5151   \\
 \hline

A4 & 0.15 & 0.60 & $2.00\times10^7$ & 2.7869  \\
 \hline

A5 & 0.30 & 0.20 & $2.50\times10^5$ & 1.9356  \\
 \hline

A6 & 0.30 & 0.40 & $2.50\times10^5$ & 2.1383   \\
 \hline

A7 & 0.30 & 0.55 & $5.00\times10^5$ & 2.2800   \\
 \hline

A8 & 0.45 & 0.35 & $2.00\times10^6$ & 2.2509  \\
 \hline

A9 & 0.45 & 0.50 & $2.50\times10^5$ & 2.1083   \\
 \hline

\end{tabular}
}
\bigskip

\noindent{
\begin{tabular}{|c|c|c|c|c|c|c|} \hline
& \multicolumn{6}{|c|}{Soft SUSY-breaking Parameters at the EW
 Scale (GeV or GeV$^2$)} \\ \hline

{\rm Pts} & $M_{1,2,3}$ &$m_{H_d}^2$ &$m_{H_u}^2$ &$m_N^2$
 & $A_{\lambda}$ &$A_{\kappa}$ \\
\hline

A1 &888.7, 2225.8, 3518.0& $-1.88\times 10^6$ &$-7.58\times 10^6$
&$-1.55\times 10^4$  & -5.0 & 0.7\\
\hline

A2 &1087.3, 2768.7, 4076.2 &$-5.61\times
10^6$ &  $-1.04\times 10^7$ & $-2.24\times10^4$ &-34.0& 1.0 \\
\hline

A3 &3561.3, 9274.7, 12311.1 &$-9.01\times
10^7$ &  $-1.18\times 10^8$ & $-2.07\times10^5$ &-268.9& 5.0 \\
\hline

A4 &3792.8, 10085.2, 12069.8&$-8.80\times
10^7$ &  $-1.15\times 10^8$ & $-3.55\times10^5$ &-241.4 & 7.3 \\
\hline

A5 & 1176.0, 2881.1, 4978.1 &$-1.71\times
10^5$ &  $-1.71\times 10^7$ & $-9.29\times10^4$ &7.7& 4.1 \\
\hline

A6 &906.6, 2276.4, 3560.4 &$-2.63\times
10^6$ &  $-7.72\times 10^6$ & $-5.66\times10^4$ &-13.3& 2.8 \\
\hline

A7 &1055.3, 2689.6, 3943.7 &$-4.27\times
10^6$ &  $-9.83\times 10^6$ & $-8.27\times10^4$ &-24.0& 3.8 \\
\hline

A8 &2070.8, 5262.9, 7810.5 &$3.62\times
10^7$ &  $-5.04\times 10^7$ & $-1.93\times10^6$ &219.1& 37.9 \\
\hline

A9 &898.7, 2248.9, 3567.5 &$3.14\times 10^6$ & $-8.05\times 10^6$
& $-2.20\times10^5$ & 45.6 & 8.3
\\ \hline

\end{tabular}

 }

\bigskip
\noindent{
\begin{tabular}{|c|c|c|c|c|c|} \hline
 & \multicolumn{5}{|c|}{Output Parameters} \\ \hline
{\rm Pts}  &$h_t$, $h_b$& $\Lambda_q$ (GeV) & ${\rm tan}\beta$& $\mu$ (GeV) & $B_\mu$ (GeV$^2$)  \\
 \hline

A1  & 0.949, 0.753 & $3.90\times 10^5$ & 43.57 & 173.8 & $1.41\times 10^4$  \\
 \hline

A2  & 0.948, 0.833 & $4.52\times 10^5$ & 48.44 & 105.1 & $7.38 \times 10^3$  \\
 \hline

A3  & 0.948, 0.880 & $1.37\times 10^6$ & 51.05 & 121.8 & $6.55 \times 10^3$  \\
 \hline

A4 & 0.948, 0.882 & $1.34\times 10^6$ & 52.41 & 106.0 & $1.92\times 10^4$  \\
 \hline

A5  & 0.949, 0.637 & $5.46\times 10^5$ & 36.93 & 321.8 & $7.11\times 10^4$  \\
 \hline

A6 & 0.948, 0.780 & $3.95\times 10^5$ & 45.30 & 124.2 & $1.88\times 10^4$  \\
 \hline

A7  & 0.948, 0.809 & $4.38\times 10^5$ & 46.89 & 109.9 & $1.94\times 10^4$  \\
 \hline

A8  & 0.950, 0.307 & $8.68\times 10^5$ & 17.80 & 1276.6 & $1.54\times 10^6$  \\
 \hline

A9 & 0.949, 0.533 & $2.50\times 10^5$ & 30.87 & 296.7 & $1.11\times 10^5$  \\
 \hline

\end{tabular}
}

\end{center}

\end{table}

\begin{table}
\caption{Mass spectrum of particles and superparticles in the
low-scale general gauge mediation.} \label{table2}
\bigskip

\begin{center}

\noindent{
\begin{tabular}{|c|c|c|c|c|} \hline
& \multicolumn{4}{|c|}{Particle Masses (TeV)}\\ \hline

{\rm Pts}   & $m_{\tilde g}$ & $m_{\tilde t_{1,2}} $ &
 $m_{\tilde b_{1,2}} $& $m_{\tilde \tau_{1,2}} $
 \\ \hline

A1  & 3.44 &5.55,  6.36 & 5.86,  6.35 &  0.80, 2.84
 \\ \hline

A2  & 3.95 &  6.37, 7.36 &
 6.60, 7.35 &  0.89, 3.53
 \\ \hline

A3  & 11.17 &  18.63, 22.24 &
 19.08,  22.24 & 2.27,  11.90
  \\ \hline

A4  & 10.98  &  17.78, 22.18 &
 18.26,  22.18 &  1.67, 12.98
 \\ \hline

A5  & 4.76 &  7.89, 8.98 &
  8.54, 8.97 &  1.14, 3.69
 \\ \hline

A6  & 3.48 & 5.62,  6.42 &
 5.89,  6.42 &  0.80, 2.90
 \\ \hline

A7  & 3.83 & 6.16,  7.14 &
  6.43, 7.14&  0.88, 3.43
 \\ \hline

A8  & 7.30 &  12.01, 14.72 &
  14.15, 14.72&  2.33, 6.85
 \\ \hline

A9  & 3.49 &  5.63, 6.57 &
 6.23,  6.57 &  0.92, 2.88
  \\ \hline
\end{tabular}
}

\bigskip

\noindent{
\begin{tabular}{|c|c|c|c|c|} \hline
& \multicolumn{4}{|c|}{Particle Masses (GeV)}\\ \hline

{\rm Pts} &  $m_{\chi_1^c} $ & $m_{\chi_1^0} $ & $m_{h_{1,2,3}}$ &
$m_{a_{1,2}}$
\\ \hline

A1  & $173.4$ & $155.8$  & 118.3, 187.3,  1751.6& 15.7,  1751.6 \\
\hline

A2  & $105.0$ & $103.7$  & 136.6, 211.1, 1616.8  &  20.3, 1616.8 \\
\hline

A3   & $121.7$ & $121.7$  & 152.6, 644.3,  3544.8& 71.3, 3544.8
\\ \hline

A4   & $106.0$ & $105.8$  &  152.4, 843.6,  3564.4 & 98.0, 3564.3  \\
\hline

A5   & 321.4 & 311.1  & 117.4, 433.3,  2825.2&  53.8, 2825.1 \\
\hline

A6   & $123.9$ & $119.8$  &  133.1, 331.2, 1656.8&  43.3, 1656.7
\\ \hline

A7   & $109.7$ & $107.1$  &  137.5, 401.8, 1754.8&  54.3, 1754.6
\\ \hline

A8   & $1276.2$ & $1272.4$  &  116.2, 1973.2, 6596.7 & 337.4,
6596.6
\\ \hline

A9   & $296.1$ & $289.8$  &  121.9, 659.6, 2430.1 & 96.1, 2429.9
\\ \hline
\end{tabular}
}

\end{center}

\end{table}

\begin{table}
\caption{Parameters of the intermediate-scale general gauge
mediation.} \label{table3}
\bigskip

\begin{center}

\noindent{
\begin{tabular}{|c|c|c|c|c|} \hline
 & \multicolumn{4}{|c|}{Input Parameters} \\ \hline
{\rm Pts} & $\lambda(\Lambda_{EW})$ & $\kappa(\Lambda_{EW})$ & $\Lambda_M$ (GeV) & $\eta$   \\
 \hline

B1 & 0.15 & 0.075 & $1.00\times10^{11}$ & 4.180  \\
 \hline

B2 & 0.15 & 0.15 & $1.00\times10^{11}$ & 4.512  \\
 \hline

B3 & 0.15 & 0.40 & $1.00\times10^{11}$ & 4.292  \\
 \hline

B4 & 0.15 & 0.60 & $1.00\times10^{11}$ & 4.126   \\
 \hline

B5 & 0.30 & 0.20 & $1.00\times10^{11}$ & 3.981   \\
 \hline

B6 & 0.30 & 0.40 & $1.00\times10^{11}$ & 4.360   \\
 \hline

B7 & 0.30 & 0.55 & $1.00\times10^{11}$ & 4.620  \\
 \hline

B8 & 0.45 & 0.35 & $1.00\times10^{11}$ & 4.019   \\
 \hline

B9 & 0.45 & 0.50 & $1.00\times10^{11}$ & 4.542  \\
 \hline

\end{tabular}
}
\bigskip

\noindent{
\begin{tabular}{|c|c|c|c|c|c|c|} \hline
& \multicolumn{6}{|c|}{Soft SUSY-breaking Parameters at the EW
 Scale (GeV or GeV$^2$)} \\ \hline

{\rm Pts} & $M_{1,2,3}$ &$m_{H_d}^2$ &$m_{H_u}^2$ &$m_N^2$
 & $A_{\lambda}$ &$A_{\kappa}$ \\
\hline

B1 &781.9, 2225.5, 1762.6 & $7.98\times 10^6$ &$-2.23\times 10^6$
&$-1.42\times 10^5$  & 379.6 & 11.6\\
\hline

B2 &443.3, 1274.9, 935.4 &$1.82\times
10^6$ &  $-4.37\times 10^5$ & $-4.41\times10^4$ &177.8 & 6.2 \\
\hline

B3 &  1177.2, 3362.9, 2593.9 &$2.41\times
10^6$ &  $-4.34\times 10^6$ & $-2.19\times10^5$ &215.8 & 13.1 \\
\hline

B4 & 2138.5, 6076.2, 4875.4&$3.85\times
10^6$ &  $-1.76\times 10^7$ & $-4.93 \times10^5$ &229.8& 19.5 \\
\hline

B5 & 1360.7, 3846.7, 3199.0 & $2.63\times
10^7$ &  $-9.56\times 10^6$ & $-1.66\times10^6$ &649.5& 80.0 \\
\hline

B6 &762.0, 2181.3, 1656.3& $7.41\times
10^6$ &  $-1.89\times 10^6$ & $-5.00\times10^5$ &357.5& 42.7 \\
\hline

B7 &475.6, 1371.9, 983.1& $2.46\times 10^6$ & $-5.17\times 10^5$ &
$-1.73\times10^5$ &205.9 & 24.4 \\ \hline

B8 &1678.7, 4752.2, 3914.6 & $3.95\times
10^7$ &  $-1.66\times 10^7$ & $-5.37\times10^6$ &730.5& 214.8 \\
\hline

B9 & 747.1, 2150.0,1567.1 & $7.86\times 10^6$ & $-1.96\times 10^6$
& $-1.04\times10^5$ &356.4& 92.8
\\ \hline

\end{tabular}
}

\bigskip

\noindent{
\begin{tabular}{|c|c|c|c|c|c|} \hline
 &  \multicolumn{5}{|c|}{Output Parameters} \\ \hline
{\rm Pts}  &$h_t$, $h_b$& $\Lambda_q$ (GeV) & ${\rm tan}\beta$& $\mu$ (GeV) & $B_\mu$ (GeV$^2$)  \\
 \hline

B1 & 0.950, 0.331 & $1.98\times 10^5$ & 19.11 & 541.4 & $3.51 \times 10^5$  \\
 \hline

B2  & 0.949, 0.550 & $1.05\times 10^5$ & 31.88 & 150.3 & $4.91 \times 10^4$  \\
 \hline

B3  & 0.949, 0.780 & $2.91 \times 10^5$ & 45.17 & 126.0 & $6.92 \times 10^4$  \\
 \hline

B4 & 0.948, 0.832 & $5.47\times 10^5$ & 48.15 & 126.2 & $9.22\times 10^4$  \\
 \hline

B5  & 0.953, 0.183 & $3.59\times 10^5$ & 10.57 & 1406.6 & $2.22\times 10^6$  \\
 \hline

B6  & 0.949, 0.340 & $1.86 \times 10^5$ & 19.62 & 384.5 & $3.33 \times 10^5$  \\
 \hline

B7  & 0.949, 0.465 & $1.10 \times 10^5$ & 26.97 & 163.5 & $8.23\times 10^4$  \\
 \hline

B8  & 0.957, 0.125 & $4.39\times 10^5$ & 7.16 & 2188.5 & $5.30\times 10^6$  \\
 \hline

B9  & 0.953, 0.173 & $1.76\times 10^5$ & 10.03 & 673.5 & $7.40\times 10^5$  \\
 \hline

\end{tabular}
}

\end{center}

\end{table}

\begin{table}
\caption{Mass spectrum of particles and superparticles in the
intermediate-scale general gauge mediation.} \label{table4}
\bigskip

\begin{center}

\noindent{
\begin{tabular}{|c|c|c|c|c|} \hline
& \multicolumn{4}{|c|}{Particle Masses (TeV)}\\ \hline

{\rm Pts}  &$m_{\tilde g}$ & $m_{\tilde t_{1,2}} $ &
 $m_{\tilde b_{1,2}} $& $m_{\tilde \tau_{1,2}} $
 \\ \hline

B1  & 1.82 &  1.98, 4.03 &
 3.23,  4.02 &  0.90, 3.07
 \\ \hline

B2  & 1.00 &  0.98, 2.16 &
 1.56,  2.16&  0.32, 1.74
 \\ \hline

B3  & 2.61 & 2.84, 5.58 &
  3.59, 5.58 &  1.00, 4.49
 \\ \hline

B4 & 4.72 &  5.52, 10.16 &
 6.42,  10.16&  2.11, 8.07
 \\ \hline

B5  & 3.19 &  3.69, 7.21 &
  6.04, 7.21 &  1.75, 5.34
  \\ \hline

B6  & 1.72 & 1.79,  3.86 &
 3.02,  3.85 &  0.87, 3.01
 \\ \hline

B7  & 1.05 &  1.00, 2.33 &
  1.71, 2.32 &  0.45, 1.88
\\ \hline

B8  & 3.86 &  4.42, 8.87 &
 7.44,  8.87 &  2.20, 6.60
  \\ \hline

B9  & 1.63 &  1.60, 3.76 &
  2.96, 3.75&  0.97, 2.98
  \\ \hline
\end{tabular}
}

\bigskip

\noindent{
\begin{tabular}{|c|c|c|c|c|} \hline
& \multicolumn{4}{|c|}{Particle Masses (GeV)}\\ \hline

{\rm Pts}
 & $m_{\chi_1^c} $ & $m_{\chi_1^0} $ & $m_{h_{1,2,3}}$ & $m_{a_{1,2}}$  \\ \hline

B1
  & $540.3$ & $520.5$  & 121.4, 536.5, 2931.9 &  97.1, 2931.9 \\ \hline

B2
  & $149.4$ & $144.7$  &  121.1, 297.6, 1416.9& 54.0,  1416.9 \\ \hline

B3
  & $125.9$ & $124.4$  & 135.3, 663.8,  2367.7&  115.8, 2367.6 \\ \hline

B4
  & $126.1$ & $125.4$  &  142.2, 997.6, 3227.7&  172.5, 3227.6 \\ \hline

B5
  & 1405.5 & 1342.6  & 120.6, 1843.4, 5383.5 & 473.7,  5383.4 \\ \hline

B6
  & $383.7$ & $380.6$  & 126.1, 1009.0,  2803.2&  192.7,  2136.1  \\ \hline

B7
  & $162.6$ & $159.3$  & 122.0, 590.0,  1623.6 &  150.5, 1623.3  \\ \hline

B8
  & $2187.3$ & $1676.5$  & 117.7, 3331.0,  6768.7 &  1045.5, 6768.5 \\ \hline

B9
  & $671.8$ & $658.9$  &  118.6, 1464.2, 2911.0 &  457.8, 2910.5 \\ \hline
\end{tabular}
}

\end{center}

\end{table}

\begin{table}
\caption{Parameters of the high-scale general gauge mediation.}
\label{table5}
\bigskip

\begin{center}

\noindent{
\begin{tabular}{|c|c|c|c|c|} \hline
 & \multicolumn{4}{|c|}{Input Parameters} \\ \hline
{\rm Pts} & $\lambda(\Lambda_{EW})$ & $\kappa(\Lambda_{EW})$ & $\Lambda_M$ (GeV) & $\eta$   \\
 \hline

C1 & 0.15 & 0.075 & $1.00\times10^{15}$ & 4.695  \\
 \hline

C2 & 0.15 & 0.15 & $1.00\times10^{15}$ & 4.980   \\
 \hline

C3 & 0.15 & 0.40 & $1.00\times10^{15}$ & 5.060  \\
 \hline

C4 & 0.15 & 0.60 & $1.00\times10^{15}$ & 4.930   \\
 \hline

C5 & 0.30 & 0.20 & $1.00\times10^{15}$ & 4.639  \\
 \hline

C6 & 0.30 & 0.40 & $1.00\times10^{15}$ & 5.110  \\
 \hline

C7 & 0.30 & 0.55 & $1.00\times10^{15}$ & 5.240   \\
 \hline

C8 & 0.45 & 0.35 & $1.00 \times10^{15}$ & 4.755   \\
 \hline

C9 & 0.45 & 0.50 & $1.00\times10^{15}$ & 5.560   \\
 \hline

\end{tabular}
}

\bigskip

\noindent{
\begin{tabular}{|c|c|c|c|c|c|c|} \hline
& \multicolumn{6}{|c|}{Soft SUSY-breaking Parameters at the EW
 Scale (GeV or GeV$^2$)} \\ \hline

{\rm Pts} & $M_{1,2,3}$ &$m_{H_d}^2$ &$m_{H_u}^2$ &$m_N^2$
 & $A_{\lambda}$ &$A_{\kappa}$ \\
\hline

C1 &  864.5, 2506.8, 1749.5 &$1.26\times 10^7$ &$-2.73\times 10^6$
&$-2.96\times 10^5$  & 742.4 & 32.1\\
\hline

C2 & 628.2, 1834.5, 1207.0 & $5.67\times
10^6$ &  $-9.95\times 10^5$ & $-1.54\times10^5$ &498.5& 22.8 \\
\hline

C3 & 833.5, 2438.6, 1579.2 & $3.85\times
10^6$ &  $-1.57\times 10^6$ & $-2.01\times10^5$ &392.7& 24.6 \\
\hline

C4 &1469.9, 4287.5, 2849.6 & $3.34\times
10^6$ &  $-5.76\times 10^6$ & $-3.79\times10^5$ &466.2& 33.2 \\
\hline

C5 &1103.6, 3195.32, 2257.0 & $2.07\times
10^7$ &  $-5.90\times 10^6$ & $-1.86\times10^6$ &877.1& 161.5 \\
\hline

C6 &742.4, 2174.7, 1394.5 &$9.02\times
10^6$ &  $-1.59\times 10^6$ & $-7.82\times10^5$ &608.1& 101.4 \\
\hline

C7 &705.1, 2071.2, 1295.2&$7.34\times
10^6$ &  $-1.24\times 10^6$ & $-5.69\times10^5$ &549.1& 84.3 \\
\hline

C8 &1981.9, 5755.8, 3966.3 & $6.35\times
10^7$ &  $-2.35\times 10^7$ & $-1.21\times10^7$ &1332.5& 611.7 \\
\hline

C9 & 781.5, 2310.7, 1361.8 & $9.85\times 10^6$ & $-2.04\times
10^6$ & $-1.67\times10^6$ &586.9& 222.2
\\ \hline

\end{tabular}
}

\bigskip

\noindent{
\begin{tabular}{|c|c|c|c|c|c|} \hline
 &  \multicolumn{5}{|c|}{Output Parameters} \\ \hline
{\rm Pts}  &$h_t$, $h_b$& $\Lambda_q$ (GeV) & ${\rm tan}\beta$& $\mu$ (GeV) & $B_\mu$ (GeV$^2$)  \\
 \hline

C1  & 0.951, 0.220 & $1.98\times 10^5$ & 12.63 & 792.6 & $8.99\times 10^5$  \\
 \hline

C2  & 0.949, 0.391 & $1.37\times 10^5$ & 22.64 & 285.4 & $2.23 \times 10^5$  \\
 \hline

C3 & 0.948, 0.702 & $1.79 \times 10^5$ & 40.79 & 122.3 & $8.75 \times 10^4$  \\
 \hline

C4  & 0.948, 0.794 & $3.23\times 10^5$ & 46.05 & 112.8 & $1.03 \times 10^5$  \\
 \hline

C5 & 0.958, 0.124 & $2.56\times 10^5$ & 7.10 & 1524.2 & $2.87\times 10^6$  \\
 \hline

C6 & 0.951, 0.233 & $1.58 \times 10^5$ & 13.47 & 492.5 & $6.20 \times 10^5$  \\
 \hline

C7 & 0.949, 0.342 & $1.47 \times 10^5$ & 19.86 & 306.8 & $3.38\times 10^5$  \\
 \hline

C8  & 0.967, 0.087 & $4.49\times 10^5$ & 4.99 & 3406.1 & $1.35\times 10^7$  \\
 \hline

C9  & 0.958, 0.123 & $1.54\times 10^5$ & 7.09 & 891.2 & $1.39\times 10^6$  \\
 \hline

\end{tabular}
}

\end{center}

\end{table}

\begin{table}
\caption{Mass spectrum of particles and superparticles in the
high-scale general gauge mediation.} \label{table6}
\bigskip

\begin{center}

\noindent{
\begin{tabular}{|c|c|c|c|c|} \hline
& \multicolumn{4}{|c|}{Particle Masses (TeV)}\\ \hline

{\rm Pts}    & $m_{\tilde g}$ & $m_{\tilde t_{1,2}} $ &
 $m_{\tilde b_{1,2}} $& $m_{\tilde \tau_{1,2}} $
 \\ \hline

C1  & 1.80 & 1.17, 4.37 &
  3.33, 4.37 &  1.18, 3.68
   \\ \hline

C2  & 1.27 & 0.61, 3.07 &
 2.15, 3.06& 0.68, 2.67
 \\ \hline

C3  & 1.63 &  0.68, 3.82 &
 2.08,  3.81 &  0.89, 3.43
  \\ \hline

C4  & 2.84 & 1.54,  6.61 &
 3.14,  6.61 &  2.08, 5.95
 \\ \hline

C5  & 2.29 &  1.48, 5.62 &
 4.37, 5.61 &  1.59, 4.71
 \\ \hline

C6  & 1.46 &  0.52, 3.65 & 2.64,  3.65 &  1.00, 3.19
  \\ \hline

C7  & 1.36 &  0.31, 3.41 &
  2.36, 3.40&  0.84, 3.02
  \\ \hline

C8  & 3.90 &  2.07, 9.98 &
 7.71,  9.98&  2.89, 8.48
 \\ \hline

C9  & 1.43 &  0.71, 3.76 &
  2.64, 3.76 &  1.14, 3.40
 \\ \hline
\end{tabular}
}

\bigskip

\noindent{
\begin{tabular}{|c|c|c|c|c|} \hline
& \multicolumn{4}{|c|}{Particle Masses (GeV)}\\ \hline

{\rm Pts}
 & $m_{\chi_1^c} $ & $m_{\chi_1^0} $ & $m_{h_{1,2,3}}$ & $m_{a_{1,2}}$  \\ \hline

C1
  & $791.3$ & $768.6$  & 120.7, 777.4, 3665.7 &  194.8, 3665.7  \\ \hline

C2
  & $284.6$ & $280.6$  &  122.2, 559.6, 2443.0 &  139.6, 2442.9 \\ \hline

C3
  & $122.1$ & $119.9$  & 126.8, 639.4,  2207.5 & 155.3, 2207.4  \\ \hline

C4
  & $112.7$ & $111.6$  & 134.2, 878.8, 2792.8 &   211.2, 2792.7 \\ \hline

C5
  & 1522.3 & 1101.0  & 116.6, 1972.0, 4872.1  & 698.9,  4871.9 \\ \hline

C6
  & 491.3 & 486.7  &   121.2, 1274.9, 3068.7& 446.3, 3068.4  \\ \hline

C7
  & 306.0 & 303.1  &  120.4, 1086.4, 2765.9 &  376.0, 2765.6 \\ \hline

C8
  & $3404.4$ & $1981.1$  & 120.1, 5084.2,  8909.4 & 2193.7, 8909.2  \\ \hline

C9
  & $889.0$ & $771.4$  &  117.2, 1884.6, 3306.1  & 806.7,  3305.3 \\ \hline
\end{tabular}
}

\end{center}

\end{table}

\begin{table}
\caption{Composition of light Higgs bosons ($\leq115$GeV). Here
``Re'' and ``Im'' denote the real and imaginary components of the
neutral Higgs fields, respectively. All light Higgs bosons
appearing in this paper are $CP$-odd, related to the explicitly
breaking of the global $U(1)_R$ symmetry. However, all of them can
satisfy the current experimental bounds since they are extremely
singlet-like.} \label{table7}
\bigskip

\begin{center}

\noindent{
\begin{tabular}{|c|c|c|c|c|} \hline
 \multicolumn{5}{|c|}{Composition of Light Higgs Bosons
(LHB) }\\ \hline

{\rm Pts} &  {\rm LHBs} & ${\rm Im}(H_d)$ & ${\rm Im}(H_u)$ &
${\rm Im}(N)$
\\ \hline

A1  & $a_1$ &  $-1.2\times 10^{-3}$  & $-8.8\times 10^{-4}$ & 0.999999  \\
\hline

A2  & $a_1$& $-2.0\times 10^{-3}$ & $-1.1\times 10^{-5}$ & 0.999998  \\
\hline

A3   & $a_1$ &  $-2.2\times 10^{-3}$ & $9.4\times 10^{-4}$ &
0.999997
\\ \hline

A4   & $a_1$ & $-2.1\times 10^{-3}$ & $-7.5\times 10^{-5}$ &
0.999998  \\
\hline

A5   & $a_1$ & $-2.0\times 10^{-4}$ & $3.3\times 10^{-5}$ & $>0.9999995$  \\
\hline

A6   & $a_1$ &  $1.3\times 10^{-4}$& $-1.7\times 10^{-6}$&
$>0.9999995$
\\ \hline

A7   & $a_1$ &  $1.1\times 10^{-4}$& $6.1\times 10^{-5}$&
$>0.9999995$
\\ \hline

A9   & $a_1$ & $4.6\times 10^{-3}$& $1.2\times 10^{-4}$& 0.999989
\\ \hline

B1  & $a_1$ &  $-7.0\times 10^{-5}$ & $-1.0\times 10^{-5}$ &
$>0.9999995$ \\
\hline

B2  & $a_1$ & $3.8\times 10^{-4}$ & $1.7\times 10^{-5}$ & $>0.9999995$  \\
\hline

\end{tabular}
}

\end{center}

\end{table}

\begin{table}
\caption{Composition of the lightest neutralino or the NLSP in
 the low-scale general gauge mediation.} \label{table8}
\bigskip

\begin{center}

\noindent{
\begin{tabular}{|c|c|c|c|c|c|} \hline
 \multicolumn{6}{|c|}{Composition of Lightest Neutralinos  }\\ \hline

{\rm Pts}  & $\tilde B $ & $\tilde W^0$ & $\tilde H_d$ & $\tilde
H_u$ & $\tilde N$
\\ \hline

A1  & 0.021 & -0.017 & -0.451&   -0.515&
    0.728  \\
\hline

A2  &  -0.026& 0.020& 0.678& 0.713&
   -0.173  \\
\hline

A3   & 0.009 &  -0.006 & 0.710 & -0.704 & -0.024
\\ \hline

A4   & 0.009 & -0.006 & 0.710 & -0.704 & -0.019 \\
\hline

A5   & -0.020 & 0.017 & 0.658& 0.687 & -0.308  \\
\hline

A6   & -0.030 &  0.024 & 0.671 & 0.721 & -0.172
\\ \hline

A7   & -0.027 &  0.020 & 0.679& 0.722 & -0.124
\\ \hline

A8   & -0.009& 0.009 & 0.703 & 0.706 & -0.083
\\ \hline

A9  & -0.026 &  0.022 & 0.686& 0.711& -0.152 \\
\hline

\end{tabular}
}

\end{center}

\end{table}

\newpage


\end{document}